\documentstyle[epsfig]{elsart}
\begin{document}

\newcommand{\re}{\mathop{\mathrm{Re}}}
\newcommand{\im}{\mathop{\mathrm{Im}}}
\newcommand{\D}{\mathop{\mathrm{d}}}
\newcommand{\I}{\mathop{\mathrm{i}}}
\newcommand{\E}{\mathop{\mathrm{e}}}

\thispagestyle{empty}
\begin{large}
\textbf{DEUTSCHES ELEKTRONEN-SYNCHROTRON}\\
\end{large}

DESY 03-031

March 2002

\begin{eqnarray}
\nonumber &&\cr \nonumber && \cr \nonumber &&\cr
\end{eqnarray}

\begin{center}
\begin{Large}
\textbf{A Method for Ultrashort Electron Pulse Shape-Measurement
Using Coherent Synchrotron Radiation}
\end{Large}
\begin{eqnarray}
\nonumber &&\cr \nonumber && \cr
\end{eqnarray}

\begin{large}
Gianluca Geloni
\end{large}

\textsl{\\Department of Applied Physics, Technische Universiteit
Eindhoven, \\P.O. Box 513, 5600MB Eindhoven, The Netherlands}
\begin{eqnarray}
\nonumber
\end{eqnarray}
\begin{large}
Evgeni Saldin and Evgeni Schneidmiller
\end{large}

\textsl{\\Deutsches Elektronen-Synchrotron DESY, \\Notkestrasse
85, 22607 Hamburg, Germany}
\begin{eqnarray}
\nonumber
\end{eqnarray}
\begin{large}
Mikhail Yurkov
\end{large}

\textsl{\\Particle Physics Laboratory (LSVE), Joint Institute for
Nuclear Research, \\141980 Dubna, Moscow Region, Russia}
\end{center}

\newpage

\begin{frontmatter}

\journal{}

\title{
A Method for Ultrashort Electron Pulse Shape-Measurement Using
Coherent Synchrotron Radiation}

\author[tue]{G. A.~Geloni}
\author[DESY]{E. L.~Saldin}
\author[DESY]{E. A.~Schneidmiller}
\author[DUBNA]{M. V.~Yurkov}

\address[tue]{Department of Applied Physics, Technische Universiteit
Eindhoven, The Netherlands}

\address[DESY]{Deutsches Elektronen-Synchrotron (DESY), Hamburg,
Germany}

\address[DUBNA]{Joint Institute for Nuclear Research, Dubna, 141980
Moskow region, Russia}

\begin{abstract}

In this paper we discuss a method for nondestructive measurements
of the longitudinal profile of sub-picosecond electron bunches for
X-Ray Free Electron Lasers (XFELs). The method is based on the
detection of the Coherent Synchrotron Radiation (CSR) spectrum
produced by a bunch passing a dipole magnet system. This work also
contains a systematic treatment of synchrotron radiation theory
which lies at the basis of CSR. Standard theory of synchrotron
radiation uses several approximations whose applicability limits
are often forgotten: here we present a systematic discussion about
these assumptions. Properties of coherent synchrotron radiation
from an electron moving along an arc of a circle are then derived
and discussed. We describe also an effective and practical
diagnostic technique based on the utilization of an
electromagnetic undulator to record the energy of the coherent
radiation pulse into the central cone. This measurement must be
repeated many times with different undulator resonant frequencies
in order to reconstruct the modulus of the bunch form-factor. The
retrieval of the bunch profile function from these data is
performed by means of deconvolution techniques: for the present
work we take advantage of a constrained deconvolution method. We
illustrate with numerical examples the potential of the proposed
method for electron beam diagnostics at the TESLA Test Facility
(TTF) accelerator. Here we choose, for emphasis, experiments aimed
at the measure of the strongly non-Gaussian electron bunch profile
in the TTF femtosecond-mode operation. We demonstrate that a
tandem combination of a picosecond streak camera and a CSR
spectrometer can be used to extract shape information from
electron bunches with a narrow leading peak and a long tail.

\end{abstract}

\end{frontmatter}

\clearpage

\setcounter{page}{1}

\section{Introduction}

Electron bunches with very small transverse emittance and high
peak current are needed for the operation of XFELs \cite{t,l}.
This is achieved using a two-step strategy: first generate beams
with small transverse emittance using an RF photocathode and,
second, apply longitudinal compression at high energy using a
magnetic chicane. The bunch length for XFEL applications is of
order of 100 femtoseconds. Since detailed understanding of
longitudinal dynamics in this new domain of accelerator physics is
of paramount importance for FEL performance, experiments on this
subject are planned in test facilities. The femtosecond time scale
is beyond the range of standard electronic display instrumentation
and the development of nondestructive methods for the measurement
of the longitudinal beam current distribution in such short
bunches is undoubtedly a challenging problem: in this paper we
discuss one of these methods, which is based on spectral
measurements of Coherent Synchrotron Radiation (CSR) produced by a
bunch passing a dipole magnet or an undulator.

After the compression stage, the CSR pulse can be analyzed, for
example, with a spectrometer. As we point out in Section 2, we can
decompose the coherent radiation spectrum, $P(\omega)$, into the
product of the square modulus of the bunch form factor,
$\mid\bar{F} (\omega) \mid^{2}$, and the single particle radiation
spectrum, $p(\omega)$ \cite{W}. The CSR spectrum, then, provides
information on the bunch form factor, although one has to keep in
mind that, in order to find $\mid\bar{F}(\omega)\mid^{2}$ from
$P(\omega)$, one needs the quantity $p(\omega)$.

Section 3 contains a systematic treatment of synchrotron radiation
theory. All the results presented in this section are derived from
fundamental laws of electrodynamics and the reader can follow the
whole derivation process from beginning to end. We use a synthetic
approach to present the material: simple situations are studied
first, and more complicated ones are introduced gradually.
Previous experience in synchrotron radiation theory would be
helpful but is not absolutely necessary, because all the required
material is independently derived. In this respect our paper is
reasonably self-contained.

A clear message in this work is that reexamination of dogmatic
"truths" can sometimes yield surprises. For years we were led to
believe that famous Schwinger's formulas \cite{sch} are directly
applicable to the case of synchrotron radiation from dipole magnet
and even now no attention is usually paid to the region of
applicability of these expressions. While such formulas are valid
in order to describe radiation from a dipole in the X-ray range,
their long wavelength asymptotic are not valid, in general.
Analytical study of this matter was first performed in the 80's
\cite{arc}. However, standard texts on synchrotron radiation
theory do not seem to provide a derivation of the expression for
the synchrotron radiation from an electron moving along an arc of
a circle. In Section 3 we present such a derivation, which we
believe is quite simple and instructive. Properties of coherent
synchrotron radiation from dipole magnets in the time domain are
derived and discussed too.

Standard theory of synchrotron radiation relies upon other
approximations too, and it seems interesting to pay attention to
their region of applicability. To be specific, two important
limitations are discussed. First, it is usually assumed that the
observer lies at infinite distance from the source.  Second,
people don't pay attention to the fact that, in real experimental
conditions, the radiation is seen by the detector through some
limited aperture. Thus, both finite distance effects and
diffraction effects are ignored. The aperture sizes and the
distances for which the latter assumptions are valid are so large,
in long wavelength range, as to be of limited practical interest.
We can draw at least two main conclusions from Section 3: first,
long wavelength radiation spectrum distortions can arise,
physically, from violation of the far zone assumption or from
aperture limitations (or both). Second, a CSR diagnostic method
based on the famous Schwinger's formulas (or the ones derived in
\cite{arc}) is of purely theoretical interest. In real
accelerators, the long wavelength synchrotron radiation from
bending magnet in the near zone integrates over many different
vacuum chamber pieces with widely varying aperture and $p(\omega)$
is usually a very difficult quantity to calculate with great
accuracy.

An effective and practical technique based on the spectral
properties of undulator radiation can be used to characterize the
bunch profile function.  The method we describe in Section 4 uses
an electromagnetic undulator and it is based on recording the
energy of coherent radiation pulses in the central cone. This
coherent radiation energy  turns out to be proportional, per
pulse, to the square modulus of the bunch form-factor at the
resonant frequency of the fundamental harmonic.  The measurement
must be repeated many times with different undulator resonant
frequencies (which are tuned by changing the undulator parameters)
in order to reconstruct the modulus of the bunch form-factor.

The retrieval of the bunch profile function from the modulus of
its form factor is preformed, in Section 5, by means of a
deconvolution technique. For the present work we choose a
constrained deconvolution method. This consists in finding the
best estimate of the bunch profile function, $F(t)$, for a
particularly measured form-factor modulus,
$\mid\bar{F}(\omega)\mid$, including utilization of any a priori
available information  about $F(t)$. In the end of the paper we
illustrate with a numerical example the potential of the proposed
technique for diagnostics at the TESLA Test Facility accelerator.
Here we have chosen, for emphasis, an experiment aimed at
measuring the strongly non-Gaussian electron bunch profile at TTF,
Phase 2 in femtosecond mode operation. The femtosecond mode
operation is based on the experience obtained during the operation
of the TTF FEL, Phase 1 \cite{ay} and it requires one bunch
compressor only. An electron bunch with a sharp spike at the head
is prepared with an rms width of about 20 microns and a peak
current of about one kA. This spike in the bunch generates FEL
pulses with duration below one hundred femtoseconds.  We
demonstrate that a tandem combination of a picosecond streak
camera and CSR spectrometer can be used to extract shape
information from electron bunches with a narrow leading peak and a
long tail.

\section{Physics of coherent synchrotron radiation}

To begin our consideration, let us recall some well-known aspects
of CSR. From a microscopic viewpoint, the electron beam current at
the entrance of a bending magnet is made up of moving electrons
arriving randomly at the entrance of the bending magnet:

\begin{displaymath}
J(t) = (-e)\sum^{N}_{k=1}\delta(t-t_{k}) \ ,
\end{displaymath}

\noindent where $\delta(\cdot)$ is the delta function, $(-e)$ is
the (negative) electron charge, $N$ is the number of electrons in
a bunch and $t_{k}$ is the random arrival time of the electron at
the bending magnet entrance. The electron bunch profile is
described by the profile function $F(t)$. The beam current
averaged over an ensemble of bunches can be written in the form:

\begin{displaymath}
\langle J(t)\rangle = (-e)NF(t) \ .
\end{displaymath}

\noindent The profile function for an electron beam with Gaussian
current distribution is given by:

\begin{displaymath}
F(t) = \frac{1}{\sqrt{2\pi}\sigma_{\mathrm{T}}}\exp\left(
-\frac{t^{2}}{2\sigma^{2}_{\mathrm{T}}}\right) \ .
\end{displaymath}

\noindent and the probability of arrival of an electron during the
time interval $(t,t+\D t)$ is just equal to $F(t)\D t$.

The electron beam current, $J(t)$, and its Fourier transform,
$\bar{J}(\omega)$, are connected by

\begin{displaymath}
\bar{J}(\omega) = \int\limits^{\infty}_{-\infty}\exp(\I\omega t)J(t)
\D t = (-e)\sum^{N}_{k=1}\exp(\I\omega t_{k}) \ ,
\end{displaymath}

\begin{displaymath}
J(t) = \frac{1}{2\pi}\int\limits^{\infty}_{-\infty}\exp(-\I\omega
t)\bar{J}(\omega) \D\omega = (-e)\sum^{N}_{k=1}\delta(t-t_{k}) \
\end{displaymath}

\noindent and, therefore, the average value of
$\mid\bar{J}(\omega)\mid^{2}$ can be written as:

\begin{displaymath}
\langle\bar{J}(\omega)\bar{J}^{*}(\omega)\rangle =
e^{2}N + e^{2}\sum_{k\ne n}\langle\exp(\I\omega t_{k})
\rangle\langle\exp(-\I\omega t_{n})\rangle \ .
\end{displaymath}

\noindent The expression $\langle\exp(\I\omega t_{k})\rangle$ is
nothing but the Fourier transform of the bunch profile function
$F(t)$, since:

\begin{displaymath}
\langle\exp(\I\omega t_{k})\rangle = \int\limits^{\infty}_{-\infty}
F(t_{k})\exp(\I\omega t_{k})\D t_{k} = \bar{F}(\omega) \ .
\end{displaymath}

\noindent Thus we can write:

\begin{displaymath}
\langle\mid\bar{J}(\omega)\mid^{2}\rangle = e^{2}N +
e^{2}N(N-1)\mid\bar{F}(\omega)\mid^{2} \ ,
\end{displaymath}

\noindent where the Fourier transform of the Gaussian profile
function has the form:

\begin{equation}
\bar{F}(\omega) = \exp\left(-\frac{\omega^{2}\sigma^{2}_{\mathrm{T}}}{2}
\right) \ .
\label{eq:bff}
\end{equation}

Above we described the properties of the input signal in the
frequency domain. The next step is the derivation of the spectral
function connecting the Fourier amplitudes of the output field and
the Fourier amplitudes of the input signal. We will investigate
the synchrotron radiation in the framework of a one-dimensional
model. Also, we will assume that the cross section of the particle
beam is small compared with the distance to the observer, so that
the path length difference from any point of the beam cross
section to the observer are small compared to the shortest
wavelength involved.

The component of synchrotron radiation electric field in time
domain, $E_{x,y}(t)$, and its Fourier transform,
$\bar{E}_{x,y}(\omega)$, are connected by

\begin{displaymath}
E_{x,y}(t) =
\frac{1}{2\pi}\int\limits^{\infty}_{-\infty}\bar{E}_{x,y}(\omega)
\exp(-\I\omega t)\D\omega \ ,
\end{displaymath}

\noindent and the Fourier harmonic for $\omega < 0$ is defined by
the relation $\bar{E}^{*}(\omega) = \bar{E}(-\omega)$. On the
other hand, the Fourier harmonic of the electromagnetic field  and
the Fourier harmonic of the current at the dipole magnet entrance
are connected by:

\begin{displaymath}
\bar{E}_{x,y}(\omega) = A_{x,y}(\omega)\bar{J}(\omega) \ , \quad \omega
> 0 \ ,
\end{displaymath}

\noindent where $A_{x,y}(\omega)$ is the synchrotron radiation
spectral function of the dipole. Since the radiation power is
proportional to the square of the radiation field, the averaged
total power at a certain frequency $\omega$ is given by the
expression:

\begin{equation}
P(\omega) = p(\omega)[N + N(N-1)\mid\bar{F}(\omega)\mid^{2}] \ ,
\label{eq:first}
\end{equation}

\noindent where $p(\omega)$ is the radiation power from one
electron. The first term $N$ in square bracket represents ordinary
incoherent synchrotron radiation with a power proportional to the
number of radiating particles. The second term represents coherent
synchrotron radiation. The actual coherent radiation power
spectrum depends on the particular particle distribution in the
bunch. For photon wavelengths equal and longer than the bunch
length, we expect all particles within a bunch to radiate
coherently and the intensity to be proportional to the square of
the number of $N$ of particles rather than linearly proportional
to $N$, as in the usual incoherent case. This quadratic effect can
greatly enhance the radiation since the bunch population can be
from $10^{8}$ to $10^{11}$ electrons. On the other hand the
coherent radiation power falls off rapidly for wavelengths as
short or even shorter than the rms bunch length.

A method for estimating the bunch profile function $F(t)$ is to
compute the factor $p(\omega)$, to measure the CSR power spectrum
$P(\omega)$ and to deduce the factor $\mid\bar{F}(\omega)\mid^{2}$
from equation (\ref{eq:first}). However, it is clear that the
retrievable information about the profile function will, in
general, not be complete, for it is the particular measured square
modulus of the form-factor that can be obtained, not the complex
form-factor itself.

\section{Radiation from dipole magnet}

The phenomenon of coherent synchrotron radiation has been
introduced in a conceptual way in the preceding section. As was
pointed out, the CSR power spectrum is the product of the square
modulus of bunch form-factor $\mid \bar{F}(\omega)\mid^{2}$, and
the single-particle power spectrum $p(\omega)$. Any diagnostics
method based on CSR must devote therefore attention to correctly
determinate the synchrotron radiation spectrum from one electron
$p(\omega)$, for this quantity plays a critical role in the
form-factor measurement. In the present section we will deal with
the properties of the single-particle radiation  in  time and in
frequency domain as well as with the CSR properties in time
domain. In order to give the subject a semblance of continuity, it
will be desirable to introduce considerable matter which can be
found in any of standard text on synchrotron radiation theory (see
for example \cite{W},\cite{DJ},\cite{D}). The typical textbook
treatment consists in finding the expression for the synchrotron
radiation spectrum from an electron moving in a circle. However,
no attention is usually paid to the region of applicability of the
derived expressions. For instance, the standard extension of the
theory to the case of synchrotron radiation from dipole magnet is
based on the assumption that the energy spectrum formula is
equivalent to the famous Schwinger's formula \cite{sch}. While
this formula is valid for the X-ray range, it does not provide a
satisfactory description in the long wavelength asymptotic
\cite{arc}. Moreover Scwinger's formulas are found in the limit of
an infinite observer distance and under the approximation of no
limiting aperture through which the radiation is collected. Here
we pay particular attention to the region of applicability of
Schwinger's formulas; specifically, we derive expressions for the
time and frequency dependence of the electromagnetic radiation
produced by an electron moving along an arc of a circle, and we
investigate how the CSR field pulse in the time domain is modified
in more realistic situations.

\subsection{Radiation field in the time domain}

\begin{figure}[tb]
\begin{center}
\epsfig{file=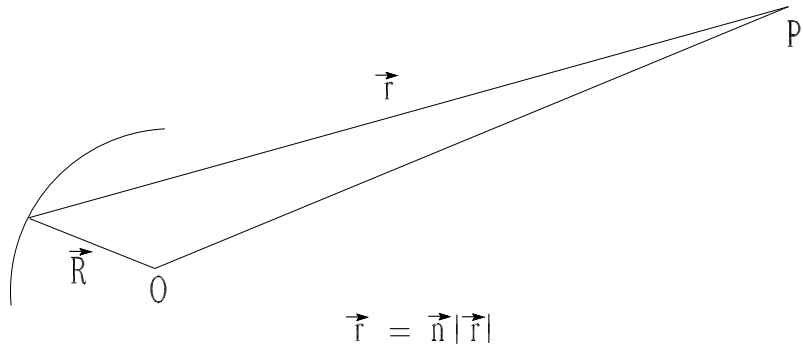,width=0.8\textwidth}
\end{center}
\caption{ Geometry for synchrotron radiation production from a
bending magnet} \label{fig:dsm1}
\end{figure}

\begin{figure}[tb]
\begin{center}
\epsfig{file=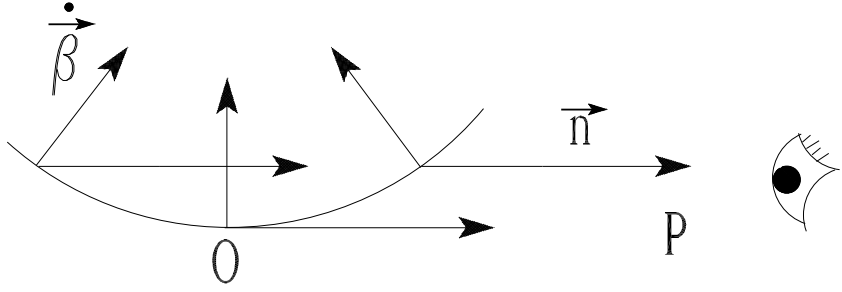,width=0.8\textwidth}
\end{center}
\caption{Synchrotron radiation production from circular motion.
Directions of acceleration vector as seen by an observer at $P$, drawn
in plane view} \label{fig:dsm9} \end{figure}

\begin{figure}[tb]
\begin{center}
\epsfig{file=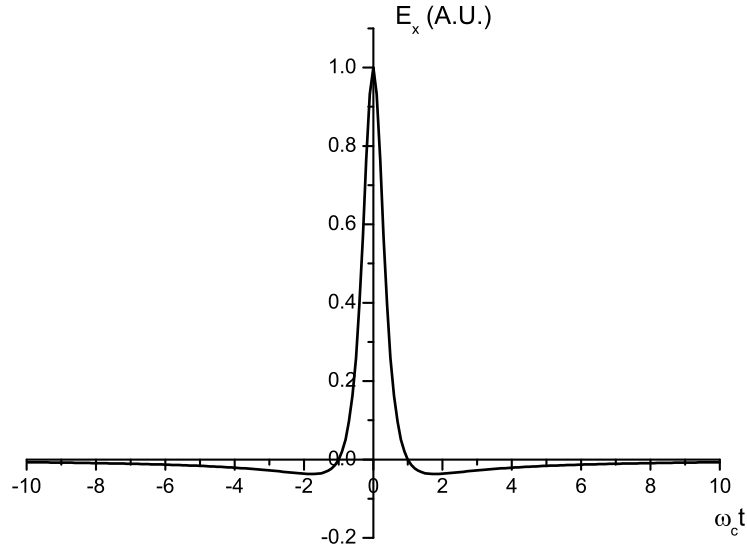,width=0.8\textwidth}
\end{center}
\caption{Time variation of a synchrotron radiation pulse generated
by a highly relativistic electron moving in a circle as seen by an
observer in the orbital plane} \label{fig:srap}
\end{figure}

\begin{figure}[tb]
\begin{center}
\epsfig{file=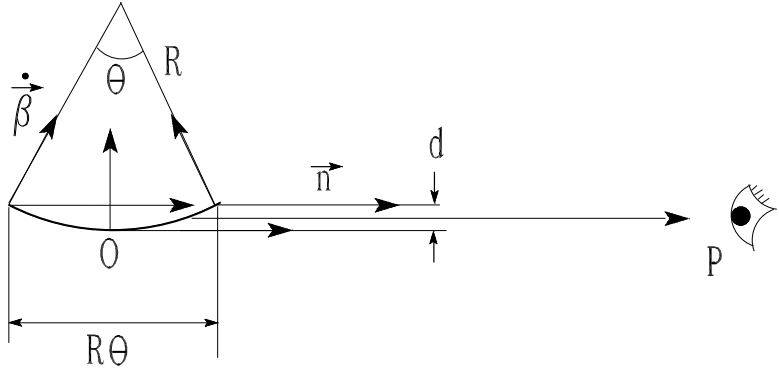,width=0.8\textwidth}
\end{center}
\caption{
Synchrotron radiation production from an arc of a circle.
Directions of acceleration vector as seen by an observer at $P$, drawn
in plane view}
\label{fig:dsm11}
\end{figure}

\begin{figure}[tb]
\begin{center}
\epsfig{file=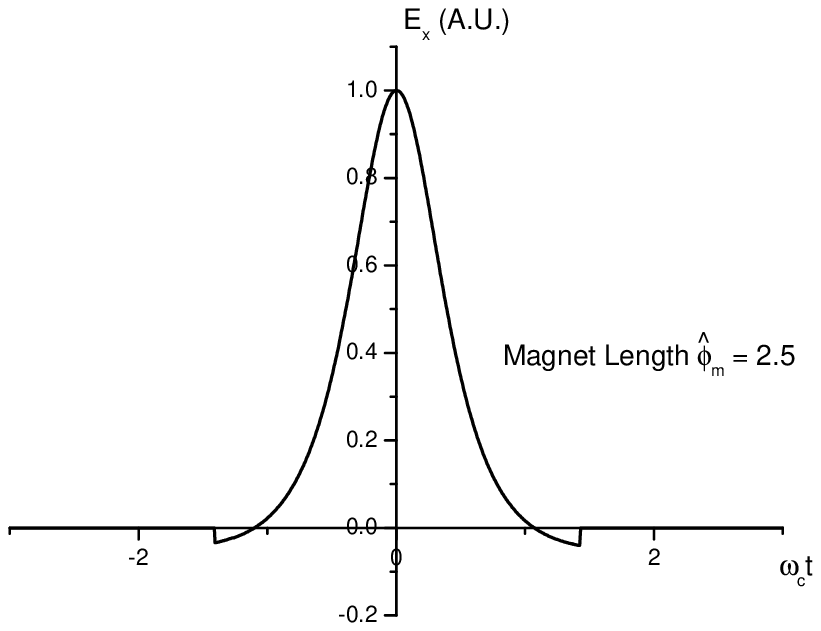,width=0.8\textwidth}
\end{center}
\caption{ Time variation of a synchrotron radiation pulse
generated by a highly relativistic electron moving along an arc of
a circle. The normalized bending angle is $\hat{\phi}_{\mathrm{m}}
= \gamma\phi_{\mathrm{m}} = 2.5$ } \label{fig:sh25a}
\end{figure}

\begin{figure}[tb]
\begin{center}
\epsfig{file=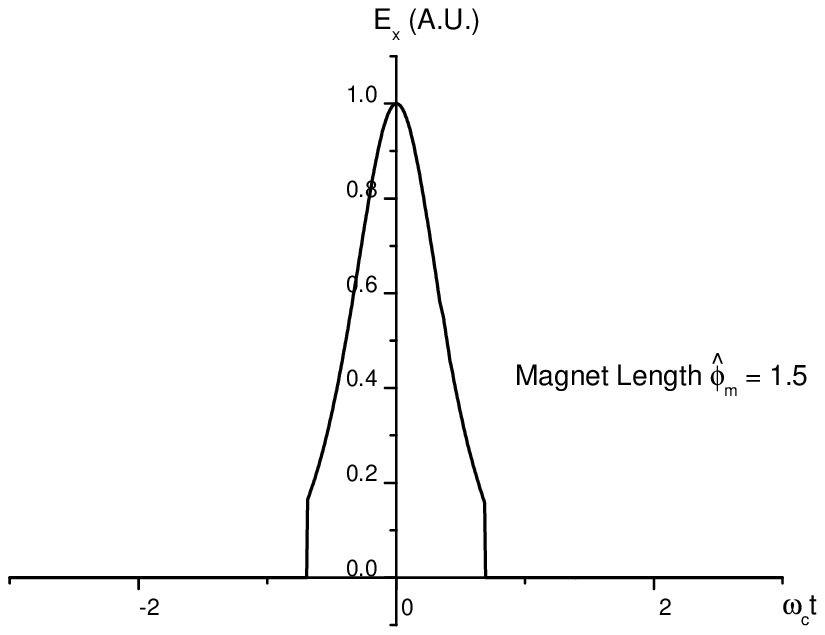,width=0.8\textwidth}
\end{center}
\caption{ Time variation of a synchrotron radiation pulse
generated by a highly relativistic electron moving along an arc of
a circle. The normalized bending angle is $\hat{\phi}_{\mathrm{m}}
= \gamma\phi_{\mathrm{m}} = 1.5$} \label{fig:sh15}
\end{figure}

\begin{figure}[tb]
\begin{center}
\epsfig{file=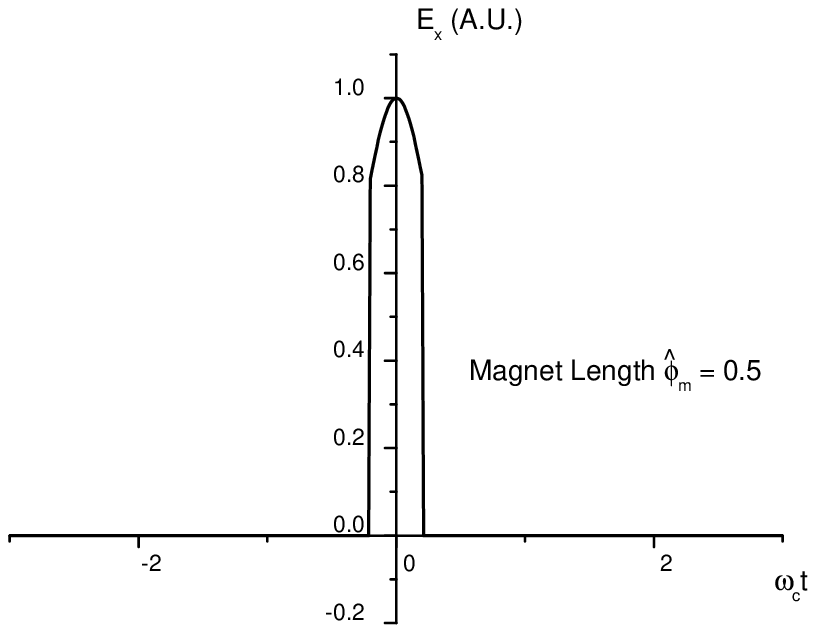,width=0.8\textwidth}
\end{center}
\caption{Time variation of a synchrotron radiation pulse from a
short bending magnet. The normalized bending angle is
$\hat{\phi}_{\mathrm{m}} = \gamma\phi_{\mathrm{m}} = 0.5$ }
\label{fig:sh05}
\end{figure}

Let us start reviewing how a single electron radiates as it moves
in a dipole magnet. The electron is subjected to an acceleration,
$\dot{\vec{v}} = c\dot{\vec{\beta}}$, of magnitude
$c^{2}\beta^{2}/R$ so that a distant observer detects the electric
and magnetic fields in the form of electromagnetic waves from the
radiating electron. Fig.~\ref{fig:dsm1} shows the relationship
between the observer at a fixed point $P$, whose coordinates are
$(\vec{r}_{0},t)$, and the radiating electron at
$(\vec{R},t^{\prime})$, $t^{\prime}$ being the emission (or
retarded) time. The fundamental laws of electrodynamics tell that
the electric field of a charge $(-e)$ moving along an arbitrary
trajectory is given by the Lienard-Wiechert formula

\begin{displaymath}
\vec{E}(t) = \frac{(-e)}{\mid\vec{r}\mid^{2}}\frac{1}{\gamma^{2}}
\left[\frac{(\vec{n} - \vec{\beta})}{(1 -
\vec{n}\cdot\vec{\beta})^{3}}\right]_{\mathrm{r}} +
\frac{(-e)}{c\mid\vec{r}\mid} \left[\frac{\vec{n}\times[(\vec{n} -
\vec{\beta})\times\dot{\vec{\beta}}]}{(1 -
\vec{n}\cdot\vec{\beta})^{3}}\right]_{\mathrm{r}} \ ,
\end{displaymath}

\noindent where $\vec{n}$ is a unit vector along the line from the
point at which the radiation is emitted at the emission time to
the observation point at the observation time, and we understand
that the quantity in brackets must be evaluated at the retarded
time $t^{\prime} = t - \frac{1}{c} \mid\vec{r}(t^{\prime})\mid$.
The latter equation consists of two distinct parts. The first is
inversely proportional to the square of the distance between
radiation source and observer, depends only on the charge velocity
and is known as velocity or Coulomb field. The second is inversely
proportional to the distance from the charge, depends also on the
charge acceleration and it is known as acceleration or radiation
field. At large distances from the moving electron, the
acceleration-related term dominates, and it is usually associated
to the electromagnetic radiation of the charge. The region of
space where the radiation field dominates is called a far (or
wave) zone and the radiative electric field in the far zone is
given by the formula

\begin{equation}
\vec{E}_{\mathrm{r}}(t) =
\frac{(-e)}{c\mid\vec{r}_{0}\mid}
\left[\frac{\vec{n}\times[(\vec{n} -
\vec{\beta})\times\dot{\vec{\beta}}]}{(1 -
\vec{n}\cdot\vec{\beta})^{3}}\right]_{\mathrm{r}} \ .
\label{eq:lwf}
\end{equation}

\noindent We can use (\ref{eq:lwf}) to look at all kinds of
interesting problems. This is a complicated expression, but it is
easy enough to be used in a computer calculation which can be
further visualized as a geometrical picture. Such geometrical
pictures will give us a good qualitative description of the
situation. Usual theory of synchrotron radiation is based on the
assumption that the electron is moving on a circle and radiation
is observed from the whole circular trajectory, so that in each
cycle we get a sharp pulse of electric field. A far-field
computation of the predicted time dependence of synchrotron
radiation for circular motion is presented in Fig.~\ref{fig:srap}.
The horizontal component of electric field plotted versus the
normalized variable $\omega_{\mathrm{c}}t$, where
$\omega_{\mathrm{c}} = 3\gamma^{3}c/(2R)$ is the critical
frequency of synchrotron radiation. The field in the orbital plane
has a zero around $t = \omega_{\mathrm{c}}^{-1}$. Numerically from
Fig. \ref{fig:srap} one obtains

\begin{equation}
\int\limits^{\infty}_{-\infty}\vec{E}_{\mathrm{r}}\D t = 0 \ .
\label{eq:z}
\end{equation}

\noindent It is interesting to stress the fact that equation
(\ref{eq:z}) is strictly related to the well-known result that a
uniformly charged ring does not radiate. In fact, starting from
(\ref{eq:z}) one can show that a system of $N$ identical
equidistant charges $(-e)$ moving with constant velocity $v$ along
a circle does not radiate in the limit for $N \to \infty$ and
$(-e)N = {\mathrm{const}}$, and the electric and the magnetic
fields of the system are the usual static values.

\begin{figure}[tb]
\begin{center}
\epsfig{file=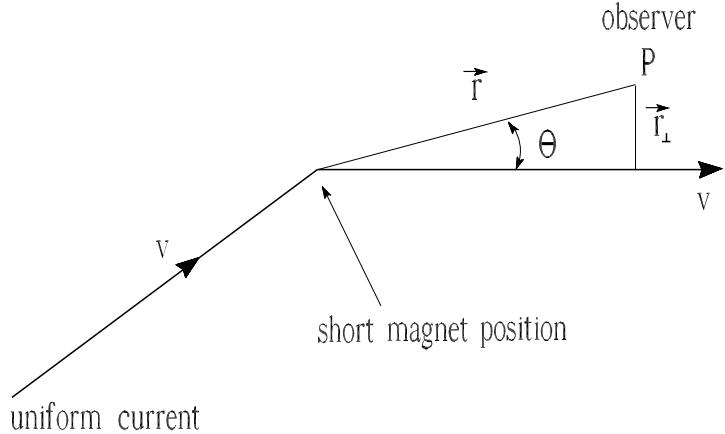,width=0.8\textwidth}
\end{center}
\caption{ Geometry for an infinite uniform current progressing
through an arc of a circle and relative position of an observer}
\label{fig:dsm22}
\end{figure}

It is important to realize that (\ref{eq:z}) is valid only when
the electron is moving in a circle. But we now want to study
synchrotron radiation from an electron moving along the arc of a
circle (see Fig. \ref{fig:dsm11}). By inspecting Fig.  \ref{fig:sh25a}
- \ref{fig:sh05}, one can see that when the electron moves in arcs of
circles with different angular extensions $\phi_m$, the time-average of
the electric field is nonzero.

Let us consider for a moment (in parallel with what has
been done in the case of a circle) the case of infinitely long electron
bunch with the homogeneous linear density $\lambda_{0}$ and current $I
= - ev\lambda_{0}$.  A current circuit consists of
the arc and semi-infinite straight lines (see Fig. \ref{fig:dsm22}).
The angle between the straight lines is equal to the bending angle
$\phi_{\mathrm{m}}$ of the magnet.
The fact that the average electric field
from a single particle is different from zero means that the
acceleration field from our infinite circuit must be different
from zero too. Then we have deal with an intriguing paradox, since
it is a well known result that an uniform electron current does
not radiate, not only in the case of circular motion, but
independently from the trajectory \footnote{It can be proved that the
radiative interaction force in the longitudinal direction (parallel, at
any time, to the velocity vector by definition) is equal to zero at any
point of such circuit}.

It is possible to explain this contradiction in very simple terms
as follows.  First of all we know that the velocity (electric)
field from a line current (including our case) is proportional to
$1/\mid\vec{r}_{\perp}\mid$, where $\mid\vec{r}_{\perp}\mid$ is
the distance of the observer from the line charge. Second, in our
case, the acceleration part of the electric field is proportional
to $1/\mid\vec{r}\mid$, where $\mid\vec{r}\mid$ is the distance of
the observer from the magnet (since the acceleration field sources
are strictly limited to the particles in the magnet only). The
situation is depicted in Fig. \ref{fig:dsm22}: the ratio
$\mid\vec{r}_{\perp}\mid/\mid\vec{r}\mid$ is finite for any
position of observer therefore there is no region in space where
the acceleration field dominates. This last observation solves our
paradox since, in the case of infinite charge current in an arc of
a circle, we cannot talk about far zone at all, although a
non-zero acceleration field is present.

\begin{figure}[tb]
\begin{center}
\epsfig{file=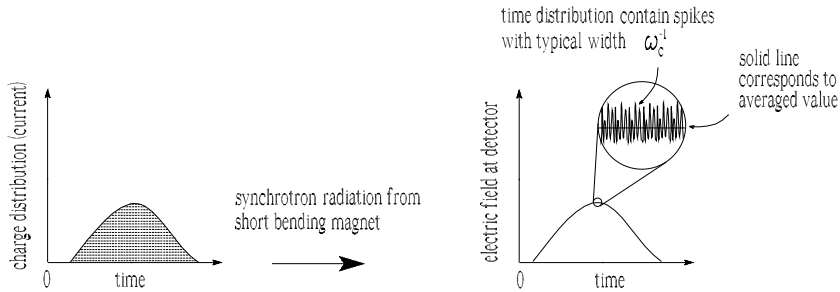,width=0.8\textwidth}
\end{center}
\caption{ Coherent synchrotron radiation production from a "short"
magnet. The time profile of the electron bunch density is linearly
encoded onto the  electric field of the radiation pulse. Spikes
correspond to spontaneous radiation, and average value of electric
field corresponds to CSR. The width of the temporal profile of the
electric field corresponds directly to the electron bunch length,
and the shape of the temporal profile is proportional to the
longitudinal distribution within the bunch} \label{fig:dsm5}
\end{figure}

It is interesting to discuss, at this point, the shape of the
coherent synchrotron radiation field pulse under different
trajectories. What we have been dealing with before is simply a
far field analysis of the single-particle radiation field. When a
large number of electrons move together, all the same way, the
total field will be a linear superposition of the individual
particle fields. Of course the result depends upon the
longitudinal distribution of the electrons. One aspect of the
problem that we can immediately deal with is the coherent
synchrotron radiation production from a "short" magnet, i.e.
$\phi_m \ll 1/\gamma$. We can assert that the field appears as
shown in Fig. \ref{fig:dsm5}. In fact the time-dependence of the
field has, for every electron, the shape shown in Fig.
\ref{fig:sh05} and the total field emitted by the electron bunch
is represented by a sum of these pulses, one for each radiating
electron. In the limit for "short" magnets, the reader can easily
come to the intuitive conclusion that the time profile of the
electron bunch density is linearly encoded onto the electric field
of the radiation pulse. The width of the temporal profile of the
electric field corresponds directly to the electron bunch length,
and the shape of the temporal profile is proportional to the
longitudinal bunch distribution.

We have just argued that the bunch density is linear encoded onto
the electric field as in Fig. \ref{fig:dsm5}; nevertheless, in
Fig. \ref{fig:dsm5}, we illustrate small fluctuations which occur
in the field amplitude too. The reason for this is that the
electron bunch is composed of large number of electrons, thus
fluctuations always exist in the electron beam density due to shot
noise effects. For any synchrotron radiation beam there is some
characteristic time, which determines the time scale of the random
field fluctuations. This characteristic time is called coherence
time $\tau_{\mathrm{c}}$ of synchrotron radiation and its
magnitude is of the order of the pulse duration from one electron.
The physical significance of these fluctuations is that there is a
short wavelength radiation component of the radiation in the range
of the inverse pulse duration from a single electron. Simple
physical considerations show that the energy spectrum of this hard
radiation component is order of $P(\omega) \simeq Np(\omega)$.
This explains the relation between small fluctuations of the
radiation field amplitude and spontaneous (incoherent) emission of
synchrotron radiation.

Up to this point we only talked about coherent radiation from a
"short" magnet. Ultimately we want to consider CSR production from
an arc of a circle. An easier step in this direction consists in
the analysis of the CSR time pulse from a circular motion. It is
possible, indeed, to derive a simple analytical expression for the
CSR pulse from a bunch with an arbitrary distribution of the
linear density satisfying the following condition:

\begin{equation}
\frac{R}{c\gamma^{3}}\frac{\D F(t)}{\D t} \ll
F(t) \ .
\label{eq:csr1}
\end{equation}

\noindent The latter condition simply indicates that the
characteristic length of the bunch is much larger than
$R/\gamma^{3}$. Let us express the total CSR pulse as a
superposition of single particle fields at a given position in the
far zone:

\begin{equation}
\vec{E}_{\mathrm{CSR}}(t) =
\int\limits^{\infty}_{-\infty}\vec{E}_{\mathrm{r}}
(t-\tau)NF(\tau)\D\tau \ ,
\label{eq:csr2}
\end{equation}

\noindent where we calibrated the observer time in such a way
that, when $F(\tau) = \delta(\tau)$, the single particle radiation
pulse has its maximum at $t = 0$.  To calculate the integral in
(\ref{eq:csr2}) one should take into account the property
(\ref{eq:z}) of the kernel $\vec{E}_{\mathrm{r}}(t-\tau)$ which
has been discussed above. Using (\ref{eq:z}) and (\ref{eq:csr1})
one can simplify equation (\ref{eq:csr2}) in the following way.
The integral (\ref{eq:csr2}) is written down as a sum of three
integrals

\begin{eqnarray}
& \mbox{} &
\vec{E}_{\mathrm{CSR}}(t) =
\int\limits^{t-\delta_{1}}_{-\infty}\vec{E}_{\mathrm{r}}
(t-\tau)NF(\tau)\D\tau
\nonumber\\
& \mbox{} &
+
\int\limits^{t+\delta_{2}}_{t-\delta_{1}}\vec{E}_{\mathrm{r}}
(t-\tau)NF(\tau)\D\tau +
\int\limits^{\infty}_{t+\delta_{2}}\vec{E}_{\mathrm{r}}
(t-\tau)NF(\tau)\D\tau \ ,
\label{eq:csr4}
\end{eqnarray}

\noindent where $\delta_{(1,2)}$ satisfy the following conditions:

\begin{equation}
\delta_{(1,2)} \gg \frac{R}{c\gamma^{3}} \ , \quad
\delta_{(1,2)}\frac{\D F(t)}{\D t} \ll F(t) \ .
\label{eq:csr5}
\end{equation}

\noindent The bunch profile function is, therefore, a slowly
varying function of the time and we simply take $F(\tau)$ outside
the integral sign and call it $F(t)$ when calculating the second
integral of (\ref{eq:csr4}):

\begin{equation}
\int\limits^{t+\delta_{2}}_{t-\delta_{1}}\vec{E}_{\mathrm{r}}
(t-\tau)NF(\tau)\D\tau
\simeq
NF(t)\int\limits^{t+\delta_{2}}_{t-\delta_{1}}\vec{E}_{\mathrm{r}}
(t-\tau)\D\tau \ .
\label{eq:csr6}
\end{equation}

\noindent Then we remember that the average of the electric field
over time is zero when an electron is moving in a circle. As a
result, we rewrite the integral in the form:

\begin{equation}
\int\limits^{t+\delta_{2}}_{t-\delta_{1}}\vec{E}_{\mathrm{r}}
(t-\tau)\D\tau =
- \left[
\int\limits^{t-\delta_{1}}_{-\infty}\vec{E}_{\mathrm{r}}
(t-\tau)\D\tau +
\int\limits^{\infty}_{t+\delta_{2}}\vec{E}_{\mathrm{r}}
(t-\tau)\D\tau \right] \ .
\label{eq:csr7}
\end{equation}

\noindent Taking into account (\ref{eq:csr6}) and (\ref{eq:csr7}), the
expression for CSR pulse becomes

\begin{eqnarray}
& \mbox{} &
\vec{E}_{\mathrm{CSR}}(t) =
\int\limits^{t-\delta_{1}}_{-\infty}\vec{E}_{\mathrm{r}}
(t-\tau)NF(\tau)\D\tau
-
NF(t)\int\limits^{t-\delta_{1}}_{-\infty}\vec{E}_{\mathrm{r}}
(t-\tau)\D\tau
\nonumber\\
& \mbox{} &
+
\int\limits^{\infty}_{t+\delta_{2}}\vec{E}_{\mathrm{r}}
(t-\tau)NF(\tau)\D\tau -
NF(t)\int\limits^{\infty}_{t+\delta_{2}}\vec{E}_{\mathrm{r}}
(t-\tau)\D\tau \ .
\label{eq:csr8}
\end{eqnarray}

\noindent Integrating by parts, the first pair of integrals on the
right hand side of (\ref{eq:csr8}) can be joined in a single one;
the same can be done with the second pair:

\begin{eqnarray}
& \mbox{} &
\vec{E}_{\mathrm{CSR}}(t) = -
\int\limits^{t-\delta_{1}}_{-\infty}
\Phi\left[\vec{E}_{\mathrm{r}}\right](t-\tau)N\frac{\D
F(\tau)}{\D\tau}\D\tau
\nonumber\\
& \mbox{} &
-
\int\limits^{\infty}_{t+\delta_{2}}
\Phi\left[\vec{E}_{\mathrm{r}}\right](t-\tau)N\frac{\D
F(\tau)}{\D\tau}\D\tau  \ ,
\label{eq:csr9}
\end{eqnarray}

\noindent where $\Phi\left[\vec{E}_{\mathrm{r}}\right]$ is a primitive
of $\vec{E}_{\mathrm{r}}$. What is left to do now is to evaluate a
primitive of the radiation field from one electron.

To calculate the primitive of $\vec{E}_{\mathrm{r}}$ we use
(\ref{eq:lwf}) and note that the electric field is expressed in
terms of quantities at the retarded time $t^{\prime}$. The
calculation is simplified if we use the following consideration:
since, in general $\D t/\D t^{\prime} =
(1-\vec{n}\cdot\vec{\beta})$ and

\begin{displaymath}
\frac{\D}{\D t^{\prime}}\left[
\frac{\vec{n}\times[(\vec{n}\times\vec{\beta})]}{(1 -
\vec{n}\cdot\vec{\beta})}\right] =
\left[\frac{\vec{n}\times[(\vec{n} -
\vec{\beta})\times\dot{\vec{\beta}}]}{(1 -
\vec{n}\cdot\vec{\beta})^{2}}\right] \ ,
\end{displaymath}

\noindent we also have

\begin{displaymath}
\frac{\D}{\D t}\left[
\frac{\vec{n}\times[(\vec{n}\times\vec{\beta})]}{(1 -
\vec{n}\cdot\vec{\beta})}\right] =
\left[\frac{\vec{n}\times[(\vec{n} -
\vec{\beta})\times\dot{\vec{\beta}}]}{(1 -
\vec{n}\cdot\vec{\beta})^{3}}\right] \ .
\end{displaymath}

\noindent Thus we can write $\vec{E}_{\mathrm{CSR}}(t)$ as

\begin{figure}[tb]
\begin{center}
\epsfig{file=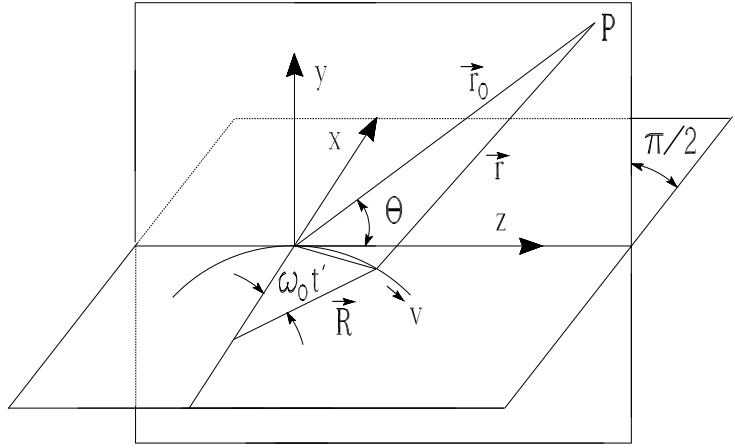,width=0.8\textwidth}
\end{center}
\caption{Geometry for synchrotron radiation from circular motion }
\label{fig:dsm19}
\end{figure}

\begin{eqnarray}
& \mbox{} & \vec{E}_{\mathrm{CSR}}(t) =
\frac{e}{c\mid\vec{r}_{0}\mid}
\int\limits^{t-\delta_{1}}_{-\infty} \left[
\frac{\vec{n}\times[(\vec{n}\times\vec{\beta})]}{(1 -
\vec{n}\cdot\vec{\beta})}\right]_{(t-\tau)} N\frac{\D
F(\tau)}{\D\tau}\D\tau
\nonumber\\
& \mbox{} & + \frac{e}{c\mid\vec{r}_{0}\mid}
\int\limits^{\infty}_{t+\delta_{2}}
\left[\frac{\vec{n}\times[(\vec{n}\times\vec{\beta})]}{(1 -
\vec{n}\cdot\vec{\beta})}\right]_{(t-\tau)} N\frac{\D
F(\tau)}{\D\tau}\D\tau  \ , \label{eq:csr10}
\end{eqnarray}

\noindent where the quantity in brackets must be evaluated at the
retarded time $t^{\prime} = (t -\tau) -
\frac{1}{c}\mid\vec{r}(t^{\prime})\mid$.

Now we would like to compute the quantities required for
(\ref{eq:csr10}). Since (\ref{eq:csr5}) holds, we can substitute
the function in brackets in both integrals on the right hand side
of (\ref{eq:csr10}) with its asymptotic behavior at $\tau \gg
R/(c\gamma^{3})$. Because the angles are very small and the
relativistic $\gamma$ factor is very large, it is very useful to
express (\ref{eq:csr10}) in a small angle approximation. The
triple vector product is calculated from Fig.  \ref{fig:dsm19}

\begin{equation}
\vec{n}\times[\vec{n}\times\vec{\beta}] =
\omega_{0}t^{\prime}\vec{e}_{x} + \theta\vec{e}_{y} \ .
\label{eq:tr}
\end{equation}

\noindent Here $\theta$ is the vertical angle, $\omega_{0} = c/R$
is the revolution frequency and $\vec{e}_{x,y}$ are unit vectors
directed along the $x$ and $y$ axis of the fixed Cartesian
coordinate system $(x,y,z)$ shown in Fig.  \ref{fig:dsm19}. The
definition of $\vec{n}$ and $\vec{R}$ can be used to compute the
scalar product in the denominator in (\ref{eq:csr10}) so that

\begin{equation}
\vec{n}\cdot\vec{\beta} = \beta\cos\theta\cos\omega_{0}t^{\prime}
\simeq \beta(1 - \theta^{2}/2)(1 - (\omega_{0}t^{\prime})^{2}/2) \ .
\label{eq:den}
\end{equation}

\noindent We assume, here, that the vertical angle is small enough
and we can leave out the cosine factor.  We can now write
$\vec{E}_{\mathrm{CSR}}(t)$ as

\begin{eqnarray}
& \mbox{} & \vec{E}_{\mathrm{CSR}}(t) =
\frac{2e}{c\mid\vec{r}_{0}\mid}
\left\{\int\limits^{t-\delta_{1}}_{-\infty}
\left[\frac{\left(\omega_{0}t^{\prime}\vec{e}_{x} +
\theta\vec{e}_{y}\right)} {(\omega_{0}t^{\prime})^{2}}
\right]_{(t-\tau)}N\frac{\D F(\tau)}{\D\tau}\D\tau \right.
\nonumber\\
& \mbox{} &
+
\left.
\int\limits^{\infty}_{t+\delta_{2}}
\left[\frac{\left(\omega_{0}t^{\prime}\vec{e}_{x}
+ \theta\vec{e}_{y}\right)}
{(\omega_{0}t^{\prime})^{2}}
\right]_{(t-\tau)}N\frac{\D
F(\tau)}{\D\tau}\D\tau\right\}  \ .
\label{eq:csr12}
\end{eqnarray}

\noindent Yet, part of the integrand in (\ref{eq:csr12}) is still
expressed as a function of $t^{\prime}$, which has to be converted in a
function of $t - \tau$ using the explicit dependence

\begin{displaymath}
t - \tau = t^{\prime} +
\frac{1}{c}\mid\vec{r}(t^{\prime})\mid = t^{\prime} -
\frac{R}{c}\cos \theta \sin \omega_{0}t^{\prime}  \ .
\end{displaymath}

\noindent Since we assume the vertical angle is very small, we may
use the replacement $\cos\theta \simeq 1$. We can therefore
approximate $t - \tau$ by

\begin{displaymath}
t^{\prime} +
\frac{1}{c}\mid\vec{r}(t^{\prime})\mid = \frac{\mid\vec{r}_{0}\mid}{c}
+ t^{\prime}(1 - \beta)
+ \frac{\omega_{0}^{2}(t^{\prime})^{3}}{6} \ .
\end{displaymath}

\noindent We conventionally fixed $\vec{E}_{\mathrm{r}}(0)$ as the
maximum value of the field (in time) and  we are interested in the
asymptotic for $\tau \gg R/(c\gamma^{3})$ only, therefore we can
simply write $t - \tau = \omega_{0}^{2}(t^{\prime})^{3}/6$ .
Solution of this equation allows us to write equation
(\ref{eq:csr12}) as

\begin{eqnarray}
& \mbox{} & \vec{E}_{\mathrm{CSR}}(t) =
\frac{2eN}{c\mid\vec{r}_{0}\mid} \left\{
\int\limits^{\infty}_{t+\delta_{2}} \frac{\left[
[6\omega_{0}(t-\tau)]^{1/3}\vec{e}_{x} +
\theta\vec{e}_{y}\right]}{[6\omega_{0}(\tau -t)]^{2/3}} \frac{\D
F(\tau)}{\D\tau}\D\tau \right.
\nonumber\\
& \mbox{} &
\left.
+
\int\limits^{t-\delta_{1}}_{-\infty}
\frac{\left[[6\omega_{0}(t- \tau)]^{1/3}\vec{e}_{x} +
\theta\vec{e}_{y}\right]}{[6\omega_{0}(t - \tau)]^{2/3}}
\frac{\D
F(\tau)}{\D\tau}\D\tau \right\} \ .
\label{eq:csr14}
\end{eqnarray}

\noindent  Limitation  (\ref{eq:csr5}) indicates that the
contribution from the integrands in the right hand side of
(\ref{eq:csr14}) are negligible in the region
$(t+\delta_{2},t-\delta_{1})$ and therefore we can rewrite
(\ref{eq:csr14}) in its final form:

\begin{equation}
\vec{E}_{\mathrm{CSR}}(t) =
\frac{2eN}{c\mid\vec{r}_{0}\mid}
\int\limits^{\infty}_{-\infty}\left[
\frac{\varepsilon(t - \tau)\vec{e}_{x}}{[6\omega_{0}\mid t
-\tau\mid]^{1/3}}
+
\frac{\theta\vec{e}_{y}}{[6\omega_{0}\mid t - \tau\mid]^{2/3}}\right]
\frac{\D
F(\tau)}{\D\tau}\D\tau \ .
\label{eq:csr15}
\end{equation}

\noindent where

\begin{displaymath}
\varepsilon(t-\tau) = 1  \qquad {\mathrm{for}}
\quad 0 < (t-\tau) < \infty \ ,
\end{displaymath}

\begin{displaymath}
\varepsilon(t-\tau) =  - 1  \qquad {\mathrm{for}} \quad
-\infty < (t-\tau) < 0 \ .
\end{displaymath}

\noindent It might be worth to remark that the ultimate reason for
using the auxiliary times $\delta_{(1,2)}$ in the derivation of
(\ref{eq:csr15}) is that they help recognizing the validity of the
asymptotic substitution, since otherwise a direct integration by
parts of (\ref{eq:csr2}) would immediately give

\begin{displaymath}
\vec{E}_{\mathrm{CSR}}(t) =  \frac{e}{c\mid\vec{r}_{0}\mid}
\int\limits^{\infty}_{-\infty} \left[
\frac{\vec{n}\times[(\vec{n}\times\vec{\beta})]}{(1 -
\vec{n}\cdot\vec{\beta})}\right]_{(t-\tau)} N\frac{\D
F(\tau)}{\D\tau}\D\tau \ .
\end{displaymath}

\noindent Under the accepted limitation on the axial gradient of the
beam current, this equation transforms to (\ref{eq:csr15}).

As an example we show how to use (\ref{eq:csr15}) in order to
calculate the CSR pulse. Let us concentrate on the CSR radiation
produced in the orbital plane. In this case $\theta = 0$ and it is
obvious that the radiation for such an observer is horizontally
polarized. To be specific, we consider an electron beam with a
Gaussian axial profile of the current density

\begin{figure}[tb]
\begin{center}
\epsfig{file=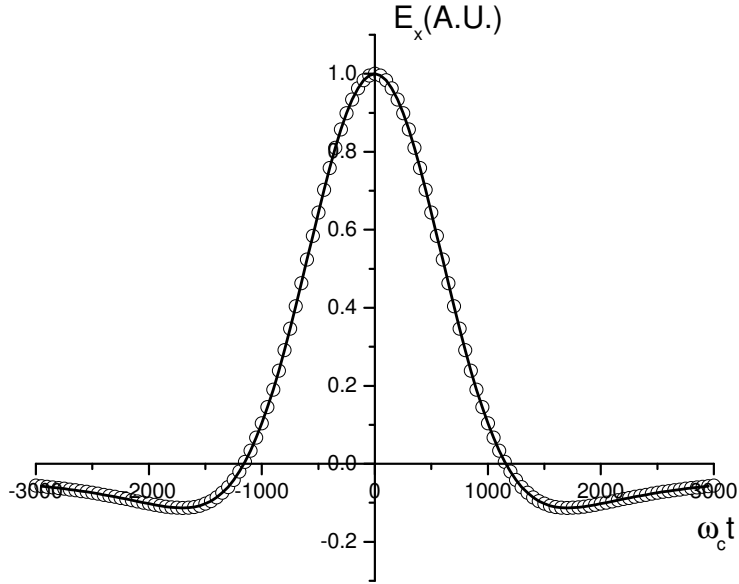,width=0.8\textwidth}
\end{center}
\caption{Time structure of a CSR pulse from a Gaussian electron
bunch moving in a circle.  Here $\theta = 0$.   Circles present
the results obtained from direct superposition of single particle
pulses, while the solid line corresponds to the shape calculated
by means of (\ref{eq:csr15}).  The rms bunch length is
$\sigma_{\mathrm{T}} = 100 \lambda_{\mathrm{c}}/c$, where
$\lambda_{\mathrm{c}} =  4\pi R/(3c\gamma^{3})$} \label{fig:csrt}
\end{figure}

\begin{displaymath}
F(t) = \frac{1}{\sqrt{2\pi}\sigma_{\mathrm{T}}}\exp\left(
-\frac{t^{2}}{2\sigma^{2}_{\mathrm{T}}}\right) \ ,
\end{displaymath}

\noindent where $\sigma_{\mathrm{T}}$ is the rms electron pulse
duration.  The rms is assumed to be large,
$\sigma_{\mathrm{T}} \gg R/(c\gamma^{3})$. When $\theta = 0$ and bunch
profile is Gaussian profile, we can write (\ref{eq:csr15}) in the form

\begin{equation}
E_{x}(t) =
\frac{2(-e)N}{(2\pi)^{1/2}6^{1/3}\sigma^{3}_{\mathrm{T}}
\omega^{1/3}_{0}c\mid\vec{r}_{0}\mid}
\int\limits^{\infty}_{-\infty} \frac{\varepsilon(t-\tau)\tau}{\mid
t - \tau\mid^{1/3}}\exp\left(
-\frac{\tau^{2}}{2\sigma^{2}_{\mathrm{T}}}\right)\D\tau \ .
\label{eq:csr14}
\end{equation}

\noindent Fig. \ref{fig:csrt} presents comparative results
obtained by means of analytical calculations (solid curve,
calculated with (\ref{eq:csr14})) and numerical results (circles,
calculated  by direct superposition of single pulses from
(\ref{eq:csr2})). It is seen from these plots that there is a very
good agreement between numerical and analytical results.

It is relevant to make some remarks about the region of
applicability of (\ref{eq:csr15}). It is important to realize that
(\ref{eq:csr15}) is valid only when the electrons are moving in a
circle and the observer is located in such a way that both the
velocity term in Lienard-Wiechert formula can be neglected and the
unit vector $\vec{n}$ can be considered constant. Another basic
assumption is that the current density changes slowly on the scale
of $R/\gamma^{3}$.  As a rule, this condition is well satisfied in
all practical problems.  It should be also mentioned that the
above expressions are good approximations only for small enough
vertical angles (even though they may be immediately generalized).
In fact we used the replacement $\cos\theta \simeq 1$ in the
retardation equation and, in practice, such an assumption is valid
for the range $\theta^{2} \ll (\sigma/R)^{2/3}$, where $\sigma$ is
the characteristic length of electron bunch.

Now we want to extend the latter results to the case of an arc of
a circle. At this point we find it convenient to impose the
following restriction\footnote{The reader can wonder why it is
necessary to describe this particular situation. The answer is
that this is a difficult subject, and the best way to study it is
to do it slowly. Although here we deal with a particular example,
all the expressions which we derive are immediately
generalizable}: we focus only on the radiation seen by an observer
lying at large distance from the sources, on the tangent to the
electrons orbit at the middle point of the magnet. In this case we
can continue to use the fixed coordinate system $(x,y,z)$ shown in
Fig. \ref{fig:dsm19}. The observation point and the vector
$\vec{n}$ are within the $(y,z)$-plane and radiation is emitted at
an angle $\theta$ with respect to the $z$-axis. Let us start
expressing the total CSR pulse as a superposition of single
particle fields at the given position in the far zone.  In the
case of an arc of a circle equation (\ref{eq:csr2}) modifies as
follows:

\begin{equation}
\vec{E}_{\mathrm{CSR}}(t) =
\int\limits^{t+T}_{t-T}\vec{E}_{\mathrm{r}}
(t-\tau)NF(\tau)\D\tau \ .
\label{eq:csr21}
\end{equation}

\noindent Here the time $T$ in the integration limits is in loco
of a window function in the integrand, in order to cut the
contributions of the single particle radiation pulse when the
electron is not in the arc. This expression contains the
observation time $T$, which should be replaced by the retarded
time $t^{\prime}_{\mathrm{e}}$. The two times are related by

\begin{displaymath}
2 T = t^{\prime}_{\mathrm{e}} +
\frac{1}{c}\mid\vec{r}(t^{\prime}_{\mathrm{e}})\mid -
\frac{1}{c}\mid \vec{r}_{0}\mid \ ,
\end{displaymath}

\noindent where $t^{\prime}_{\mathrm{e}} = \phi_{\mathrm{m}}/(
\omega_{0})$ and $\phi_{\mathrm{m}}$ is the bending magnet angular
extension. Our analysis focuses on the case of a long bending
magnet, $\gamma\phi_{\mathrm{m}} \gg 1$. Using (\ref{eq:z}) and
(\ref{eq:csr1}), the field of the CSR pulse is readily shown to be

\begin{eqnarray}
& \mbox{} &
\vec{E}_{\mathrm{CSR}}(t) =
\int\limits^{t-\delta}_{t-T}\vec{E}_{\mathrm{r}}
(t-\tau)NF(\tau)\D\tau
-
NF(t)\int\limits^{t-\delta}_{-\infty}\vec{E}_{\mathrm{r}}
(t-\tau)\D\tau
\nonumber\\
& \mbox{} &
+
\int\limits^{t+T}_{t+\delta}\vec{E}_{\mathrm{r}}
(t-\tau)NF(\tau)\D\tau -
NF(t)\int\limits^{\infty}_{t+\delta}\vec{E}_{\mathrm{r}}
(t-\tau)\D\tau \ .
\label{eq:csr81}
\end{eqnarray}

\noindent As we have already done previously, we assume that
$\delta$ satisfies condition (\ref{eq:csr5}). Adding and
subtracting suitable edge terms one can still perform integration
by parts, thus obtaining:

\begin{eqnarray}
& \mbox{} &
\vec{E}_{\mathrm{CSR}}(t) =
NF(t + T)\int\limits^{t+T}_{-\infty}\vec{E}_{\mathrm{r}}
(t-\tau)\D\tau   -
NF(t - T)\int\limits^{t-T}_{-\infty}\vec{E}_{\mathrm{r}}
(t-\tau)\D\tau
\nonumber\\
& \mbox{} &
-
\int\limits^{t-\delta}_{t-T}
\Phi\left[\vec{E}_{\mathrm{r}}\right](t-\tau)N\frac{\D
F(\tau)}{\D\tau}\D\tau  -
\int\limits^{t+T}_{t+\delta}
\Phi\left[\vec{E}_{\mathrm{r}}\right](t-\tau)N\frac{\D
F(\tau)}{\D\tau}\D\tau  \ .
\label{eq:csr91}
\end{eqnarray}

\noindent Since condition (\ref{eq:csr5}) holds for $\delta$  we
may substitute the 3rd and the 4th integral in (\ref{eq:csr91})
with a single integral in which the primitive,
$\Phi\left[\vec{E}_{\mathrm{r}}\right]$, is substituted by its
asymptotic for large values of the argument,
$\Phi\left[\vec{E}^{\mathrm{A}}_{\mathrm{r}}\right]$. Under the
assumption of a long bending magnet $(\omega_{\mathrm{c}}T \gg 1)$ the
1st and the 2nd integral can be expressed by means of the primitive
asymptotic too. Moreover, we can perform a change of variables in
all the integrals $t - \tau \to \tau$.  As a result expression
(\ref{eq:csr91}) can be written in the form:

\begin{eqnarray}
& \mbox{} &
\vec{E}_{\mathrm{CSR}}(t) = -
NF(t + T)\int\limits^{\infty}_{-T}\vec{E}^{\mathrm{A}}_{\mathrm{r}}
(\tau)\D\tau   -
NF(t - T)\int\limits^{T}_{\infty}\vec{E}^{\mathrm{A}}_{\mathrm{r}}
(\tau)\D\tau
\nonumber\\
& \mbox{} &
-
\int\limits^{T}_{-T}
\Phi\left[\vec{E}^{\mathrm{A}}_{\mathrm{r}}\right](\tau)N\frac{\D
F(t-\tau)}{\D\tau}\D\tau  \ .
\label{eq:csr92}
\end{eqnarray}

\noindent Using the ultrarelativistic approximation we can
calculate a primitive $\Phi\left[\vec{E}_{\mathrm{r}}\right]$
using (\ref{eq:tr}) and (\ref{eq:den}). Again we assume that the
vertical angle is very small so that we may use the replacement
$\cos\theta \simeq 1$. In this situation we have

\begin{displaymath}
\frac{\vec{n}\times[(\vec{n}\times\vec{\beta})]}{(1 -
\vec{n}\cdot\vec{\beta})} \simeq
\frac{\omega_{0}t^{\prime}\vec{e}_{x} + \theta\vec{e}_{y}}
{(\omega_{0}t^{\prime})^{2}/2}
\end{displaymath}

\noindent This quantity  must be evaluated at the retarded time
$t^{\prime} \simeq [6\tau/\omega^{2}_{0}]^{1/3}$. Substitution
these expressions in (\ref{eq:csr92}) gives

\begin{eqnarray}
& \mbox{} & \vec{E}_{\mathrm{CSR}}(t) =
\frac{2eN}{c\mid\vec{r}_{0}\mid} \left\{\int\limits^{T}_{-T}\left[
\frac{\varepsilon(\tau)\vec{e}_{x}}{[6\omega_{0}\mid
\tau\mid]^{1/3}} + \frac{\theta\vec{e}_{y}}{[6\omega_{0}\mid
\tau\mid]^{2/3}}\right] \frac{\D F(t-\tau)}{\D\tau}\D\tau \right.
\nonumber\\
& \mbox{} &
\left.
+
\left[F(t+T) + F(t-T)\right]\frac{ \vec{e}_{x}}{(6\omega_{0}T)^{1/3}}
\right.
\nonumber\\
& \mbox{} &
\left.
-
\left[F(t+T) -
F(t-T)\right]\frac{\theta\vec{e}_{y}}{(6\omega_{0}T)^{2/3}}\right\} \ ,
\label{eq:csr95}
\end{eqnarray}

\noindent  where $T = \phi_{\mathrm{m}}^{3}/(12\omega_{0})$.

\begin{figure}[tb]
\begin{center}
\epsfig{file=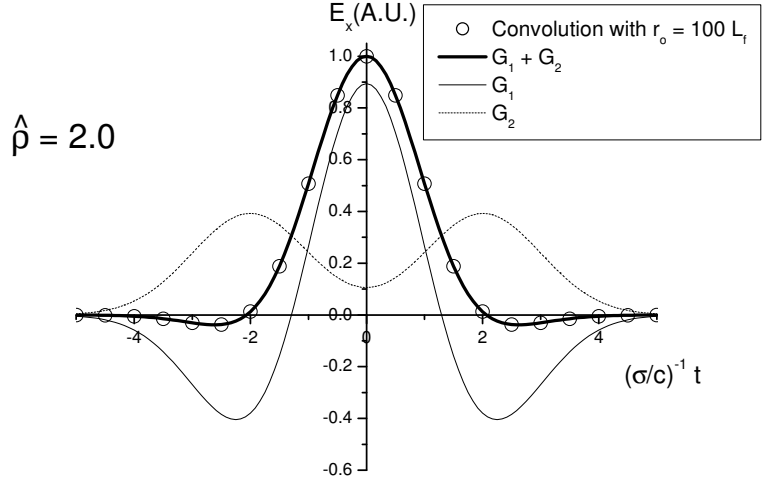,width=0.8\textwidth}
\end{center}
\caption{Time structure of the CSR pulse from a Gaussian electron
bunch moving along an arc of a circle. Here $\theta = 0, \
\hat{\rho} =
\phi^{3}_{\mathrm{m}}/(6\omega_{0}\sigma_{\mathrm{T}}) = 2$. The
physical meaning of the magnet length parameter $\hat \rho$ will
be clear after reading Section 3.3. Circles present the results
obtained from direct superposition of single particle pulses. The
solid line corresponds to the shape calculated by means of
(\ref{eq:csr95}) } \label{fig:t2}
\end{figure}

As an example of the application of this expression, consider the
situation when $\theta = 0$ and bunch profile is a Gaussian.
According to (\ref{eq:csr95}) the CSR field in this case is given
by

\begin{displaymath}
E_{x}(t) =  G_{1}(t) + G_{2}(t)
\end{displaymath}

\noindent where

\begin{displaymath}
G_{1} = \frac{2(-e)N}{(2\pi)^{1/2}6^{1/3}\sigma_{\mathrm{T}}^{3}
\omega^{1/3}_{0}c\mid\vec{r}_{0}\mid} \int\limits^{T}_{-T}
\frac{\varepsilon(\tau)(t-\tau)}{\mid \tau\mid^{1/3}}\exp\left(
-\frac{(t-\tau)^{2}}{2\sigma^{2}_{\mathrm{T}}}\right)\D\tau \ ,
\end{displaymath}

\begin{displaymath}
G_{2} =
\frac{2eN}{(2\pi)^{1/2}6^{1/3}\sigma_{\mathrm{T}}
(\omega_{0}T)^{1/3}c\mid\vec{r}_{0}\mid}
\left[
\exp\left(-\frac{(t+T)^{2}}{2\sigma^{2}_{\mathrm{T}}}\right)
+ \exp\left(-\frac{(t-T)^{2}}{2\sigma^{2}_{\mathrm{T}}}\right)
\right] \ .
\end{displaymath}

\noindent Fig. \ref{fig:t2} presents the results of calculations
with the analytical formula (\ref{eq:csr95}). Comparison with
numerical results shows a very good agreement.

\subsection{Radiation field in the frequency domain}

Let us now go back to our single particle  treatment and proceed
to the calculation of the spectrum of the electromagnetic
radiation produced by an electron during a dipole magnet pass. We
know that, in general, for any kind of motion

\begin{displaymath}
\frac{\D^{2}W}{\D\Omega\D t} = \frac{c\mid\vec{r}\mid^{2}\mid
\vec{E}(t)\mid^{2}}{4\pi} \ .
\end{displaymath}

\noindent Here $\mid\vec{E}\mid^{2}$ is the squared modulus of the
electromagnetic field vector at the observation point. The
stationary observer who is detecting the radiation emitted into
the solid angle $\D\Omega$ measures the total energy as

\begin{displaymath}
\frac{\D W}{\D\Omega} =
\frac{c\mid\vec{r_0}\mid^{2}}{4\pi}\int\limits^{\infty}_{-\infty}\mid
\vec{E}_{\mathrm{r}}(t)\mid^{2}\D t \ ,
\end{displaymath}

\noindent where we used the fact that, in the far field
approximation, the distance $\mid\vec r\mid$ is much larger than
$\mid\vec{R}\mid$ , and the zero order expansion reads
$\mid\vec{r}(t^{\prime})\mid = \mid\vec{r}_{0}\mid$ (see
Fig.~\ref{fig:dsm1}). From the discussion above, we know that the
radiation pulse is emitted over a very short period of time so
that the only finite contribution to this integral comes from
times close to $t = 0$. Extension of the integral to infinite
times is only a mathematical convenience which does not affect the
physical result.

The general method to derive the frequency spectrum is to
transform the electric field from the time domain to the frequency
domain.
Expressing the electrical field $\vec{E}_{\mathrm{r}}(t)$ by
its Fourier transform, we set

\begin{equation}
\vec{E}_{\mathrm{r}}(\omega) = \int\limits^{\infty}_{-\infty}
\vec{E}_{\mathrm{r}}(t)\exp(-\I\omega t)\D t \ .\label{eq:ft}
\end{equation}

\noindent Applying Parseval's theorem we have

\begin{displaymath}
\int\limits^{\infty}_{-\infty}\mid\vec{E}_{\mathrm{r}}(\omega)\mid^{2}
\D\omega = 2\pi\int\limits^{\infty}_{-\infty}
\mid\vec{E}_{\mathrm{r}}(t)\mid^{2}\D t
\end{displaymath}

\noindent and the total absorbed radiation energy from a single
pass is therefore

\begin{displaymath}
\frac{\D W}{\D\Omega} = \frac{c}{8\pi^{2}}\mid\vec{r_0}\mid^{2}
\int\limits^{\infty}_{-\infty}\mid\vec{E}_{\mathrm{r}}(\omega)\mid^{2}
\D\omega \ .
\end{displaymath}

\noindent Evaluating the electric field by Fourier components we
derive an expression for the spectral distribution of the
radiation energy

\begin{equation}
\frac{\D^{2}W}{\D\omega\D\Omega} = \frac{c}{4\pi^{2}}
\mid\vec{r_0}\mid^{2} \mid\vec{E}_{\mathrm{r}}(\omega)\mid^{2}
\label{eq:es}
\end{equation}

In order to calculate the Fourier transform we can use
(\ref{eq:lwf}) and note that the electric field is expressed in
terms of quantities at the retarded time $t^{\prime}$. The
calculation is simplified if we express the whole integrand in
(\ref{eq:ft}) at the retarded $t^{\prime}$ remembering $t^{\prime}
= t - \frac{1}{c}\mid\vec{r}(t^{\prime})\mid$:

\begin{displaymath}
\vec{E}_{\mathrm{r}}(\omega) =
\frac{(-e)}{c\mid\vec{r}\mid}\int\limits^{\infty}_{-\infty}
\left[\frac{\vec{n}\times[(\vec{n} -
\vec{\beta})\times\dot{\vec{\beta}}]}{(1 -
\vec{n}\cdot\vec{\beta})^{2}}\right] \exp[\I\omega(t^{\prime}+
\mid\vec{r}\mid/c)]\D t^{\prime} \ .
\end{displaymath}

\noindent Since the distance $\mid\vec{r}(t^{\prime})\mid$ is much
larger than $\mid\vec{R}\mid$ , the zero order approximation would
make $\mid\vec{r}(t^{\prime})\mid = \mid\vec{r}_{0}\mid$ (see
Fig.~\ref{fig:dsm1}). However, such approximation is not good
enough in the exponential factor and we must take the next
approximation order writing $\mid\vec{r}(t^{\prime})\mid =
\mid\vec{r}_{0}\mid - \vec{n}\cdot\vec{R}(t^{\prime})$. Therefore
we get

\begin{equation}
\vec{E}_{\mathrm{r}}(\omega) =
\frac{(-e)}{c\mid\vec{r}_{0}\mid}\int\limits^{\infty}_{-\infty}
\left[\frac{\vec{n}\times[(\vec{n} -
\vec{\beta})\times\dot{\vec{\beta}}]}{(1 -
\vec{n}\cdot\vec{\beta})^{2}}\right] \exp[\I\omega(t^{\prime}-
\vec{n}\cdot\vec{R}(t^{\prime})/c)]\D t^{\prime} \ . \label{eq:fc}
\end{equation}

\noindent We established some basic equations in the frequency
domain with which we can start calculations for individual
problems. Analytical calculations can be performed without big
difficulty in two limiting cases, namely the case of circular
motion and "short" magnet.

\subsubsection{Radiation from circular motion}

In his paper on synchrotron radiation, Schwinger gives remarkable
formulas for the radiation spectrum in the case of an electron
moving in a circle \cite{sch}. One can find textbooks telling that
Schwinger's formulas apply to the analysis of synchrotron
radiation from an electron moving along the arc of a circle too
(see for example \cite{D}).  This extension is not a physical law:
it is merely the statement of an approximation which is valid
about the entire wavelength range interesting for the ordinary
user. In this context, a critical study of the theoretical status
of Schwinger's formulas seems to be of considerable importance.

We are going to apply (\ref{eq:fc}) to our analysis of synchrotron
radiation from circular motion taking advantage of an expression
we previously used in Section 3.1: in the case of circular motion
$\vec{E}_{\mathrm{r}}(\omega)$ can be evaluated remembering that

\begin{displaymath}
\frac{\D}{\D t^{\prime}}\left[
\frac{\vec{n}\times[(\vec{n}\times\vec{\beta})]}{(1 -
\vec{n}\cdot\vec{\beta})}\right] =
\left[\frac{\vec{n}\times[(\vec{n} -
\vec{\beta})\times\dot{\vec{\beta}}]}{(1 -
\vec{n}\cdot\vec{\beta})^{2}}\right] \ ,
\end{displaymath}

\noindent Integration by parts yields

\begin{eqnarray}
& \mbox{} &
\int\limits^{\infty}_{-\infty}
\left[\frac{\vec{n}\times[(\vec{n} -
\vec{\beta})\times\dot{\vec{\beta}}]}{(1 -
\vec{n}\cdot\vec{\beta})^{2}}\right]
\exp[\I\omega(t^{\prime}- \vec{n}\vec{R}(t^{\prime})/c)]\D
t^{\prime}
\nonumber\\
& \mbox{} &
=
\left. \frac{\vec{n}\times[\vec{n}\times\vec{\beta}]}{(1 -
\vec{n}\cdot\vec{\beta})}
\exp\left[\I\omega(t^{\prime}-
\vec{n}\cdot\vec{R}(t^{\prime})/c)\right]
\right|^{t^{\prime}=\infty}_{t^{\prime}=-\infty} \nonumber\\ & \mbox{}
& -\I\omega\int\limits^{\infty}_{-\infty}
\vec{n}\times[\vec{n}\times\vec{\beta}]
\exp\left[\I\omega(t^{\prime}-
\vec{n}\cdot\vec{R}(t^{\prime})/c)\right] \D t^{\prime} \ .
\label{eqn1}
\end{eqnarray}

\noindent The reason for using integration by parts is that the
contribution from the first term (which to be evaluated at
$t^{\prime} = \pm \infty$) is zero. Thus we can write

\begin{equation}
\frac{\D^{2}W}{\D\omega\D\Omega} =
\frac{e^{2}\omega^{2}}{4\pi^{2}c}\left|\int\limits^{\infty}_{-\infty}
\vec{n}\times[\vec{n}\times\vec{\beta}]
\exp\left[\I\omega(t^{\prime}-
\vec{n}\cdot\vec{R}(t^{\prime})/c)\right] \D t^{\prime}\right|^{2} \ .
\label{eq:vf}
\end{equation}

\noindent We must emphasize, however, that this expression is not
valid in general, but only when a particle is moving in a circle.

The integrand in (\ref{eq:vf}) can be expressed in components to
simplify the integration.  If we look in the plane of the circle,
radiation does not depend upon the azimuthal angle. There is a
better way to write out the integrand in (\ref{eq:vf}) by making
use of the azimuthal symmetry, which allow one to use again the
fixed coordinate system $(x,y,z)$ shown in Fig.~\ref{fig:dsm19}.
The observation point is far away from the source point and we
focus on the radiation that is centered about the tangent to the
orbit at the source point.  The observation point $P$ and the
vector $\vec{r}$ and $\vec{n}$ are therefore within the
$(y,z)$-plane and radiation is emitted at angle $\theta$ with
respect to the $z$-axis.  Following the above discussion the
azimuthal angle is constant and set to $\phi = \pi/2$.  The
definition of $\vec{n}$ and $\vec{R}$ can be used to compute the
scalar product in the exponential term in (\ref{eq:vf}) so that

\begin{displaymath}
\vec{n}\cdot\vec{R} = R\cos\theta\sin\omega_{0}t^{\prime} \ ,
\end{displaymath}

\noindent where $\theta$ is the vertical angle and $\omega_{0} =
c/R$. Since the angles are  small and the relativistic $\gamma$
factor is  large, we can replace

\begin{displaymath}
\cos\theta \simeq 1 - \theta^{2}/2 \ , \quad
\sin\omega_{0}t^{\prime} \simeq \omega_{0}t^{\prime} - \omega^{3}_{0}
(t^{\prime})^{3}/6 \ .
\end{displaymath}

\noindent Therefore the exponential factor becomes

\begin{displaymath}
t^{\prime} - \frac{R}{c}\cos\theta\sin\omega_{0}t^{\prime}
= \frac{t^{\prime}}{2\gamma^{2}}\left(1 + \gamma^{2}\theta^{2}\right)
+ \frac{c^{2}(t^{\prime})^{3}}{6R^{2}} \ .
\end{displaymath}

\noindent The triple vector product in (\ref{eq:fc}) can be
evaluated in a similar way as

\begin{displaymath}
\vec{n}\times[\vec{n}\times\vec{\beta}]
= \omega_{0}t^{\prime}\vec{e}_{x} + \theta\vec{e}_{y} \ ,
\end{displaymath}

\noindent We can now write $\vec{E}_{\mathrm{r}}(\omega)$ as

\begin{eqnarray}
& \mbox{} & \vec{E}_{\mathrm{r}}(\omega) =
\frac{(-e)\omega}{c\mid\vec{r}\mid}
\left\{\vec{e}_{x}\left[\int\limits^{\infty}_{-\infty}
\omega_{0}t^{\prime}\exp\left(
\I\omega\left(\frac{t^{\prime}}{2\gamma^{2}}\left(1 +
\gamma^{2}\theta^{2}\right) +
\frac{c^{2}(t^{\prime})^{3}}{6R^{2}}\right)\right) \D
t^{\prime}\right] \right.
\nonumber\\
& \mbox{} &
\left.
+
\vec{e}_{y}
\left[\int\limits^{\infty}_{-\infty}
\theta\exp\left(
\I\omega\left(\frac{t^{\prime}}{2\gamma^{2}}\left(1 +
\gamma^{2}\theta^{2}\right) +
\frac{c^{2}(t^{\prime})^{3}}{6R^{2}}\right)\right) \D t^{\prime}\right]
\right\} \ .
\label{eqn2}
\end{eqnarray}

\noindent The integrals in (\ref{eqn2}) can be expressed in terms
of modified Bessel functions as first pointed out by Schwinger
\cite{sch}. For this we need a change of variables as follows:

\begin{displaymath}
\tau =
\frac{\gamma\omega_{0}t^{\prime}}{(1+\gamma^{2}\theta^{2})^{1/2}} \ ,
\quad \xi = \frac{\omega}{2\omega_{\mathrm{c}}}\left(1 +
\gamma^{2}\theta^{2}\right)^{3/2} \ ,
\end{displaymath}

\noindent where $\omega_{\mathrm{c}} = 3\gamma^3\omega_{0}/2$. With
these substitutions we get integrals of the form

\begin{displaymath}
\int\limits^{\infty}_{0}\tau\sin\left(\frac{3}{2}\xi\left(\tau
+ \frac{1}{3}\tau^{3}\right)\right)\D\tau
= \frac{1}{\sqrt{3}}K_{2/3}(\xi) \ ,
\end{displaymath}

\begin{displaymath}
\int\limits^{\infty}_{0}\cos\left(\frac{3}{2}\xi\left(\tau
+ \frac{1}{3}\tau^{3}\right)\right)\D\tau
= \frac{1}{\sqrt{3}}K_{1/3}(\xi) \ ,
\end{displaymath}

\noindent The Fourier transform of the electrical field (\ref{eqn2})
finally becomes

\begin{displaymath}
\vec{E}_{\mathrm{r}}(\omega) =
\frac{\sqrt{3}(-e)\omega}{c\mid\vec{r}_{0}\mid\omega_{\mathrm{c}}}
\gamma(1 + \gamma^{2}\theta^{2}) \left[K_{2/3}(\xi)\vec{e}_{x} +
\I\frac{\gamma\theta K_{1/3}(\xi)} {(1 +
\gamma^{2}\theta^{2})^{1/2}}\vec{e}_{y}\right] \ .
\end{displaymath}

\noindent Using (\ref{eq:es}) we get an expression for the
spectral distribution:

\begin{equation}
\frac{\D^{2}W}{\D\omega\D\Omega} =
\frac{3e^{2}\omega^{2}}{4\pi^{2}c\omega^{2}_{\mathrm{c}}}
\gamma^{2}(1 + \gamma^{2}\theta^{2})^{2}
\left[K^{2}_{2/3}(\xi) + \frac{\gamma^{2}\theta^{2}}
{1 + \gamma^{2}\theta^{2}}K^{2}_{1/3}(\xi)\right] \ .
\label{eq:esk}
\end{equation}

\noindent Quite often an observer is interested in the synchrotron
radiation energy emitted over all vertical angles. Integration of
equation (\ref{eq:esk}), over the angle $\theta$ gives the
required result. We note that the solid angle $\D\Omega =
\D\theta\D\phi$ and that radiation does not depend upon the
azimuthal angle. So we can write

\begin{equation}
\frac{\D^{2}W}{\D\omega\D\phi} =
\frac{3e^{2}\omega^{2}\gamma^{2}}{4\pi^{2}c\omega^{2}_{\mathrm{c}}}
\int\limits^{\infty}_{-\infty}
(1 + \gamma^{2}\theta^{2})^{2}
\left[ K^{2}_{2/3}(\xi) +
\frac{\gamma^{2}\theta^{2}K^{2}_{1/3}(\xi)}{1 +
\gamma^{2}\theta^{2}}\right]\D\theta \ .
\label{eq:ies} \end{equation}

\noindent The angle $\theta$ appears in (\ref{eq:ies}) in rather a
complicated way which makes it difficult to perform the
integration directly.  We shall not describe the integration
process in detail, not only because it is available in the form of
comprehensive treatises \cite{W}, but also because this is a pure
mathematical problem. If we look it up in \cite{W} we see that
(\ref{eq:ies}) can be written as

\begin{equation}
\frac{\D^{2}W}{\D\omega\D\phi} =
\frac{\sqrt{3}e^{2}\gamma}{4\pi}
\left[\frac{\omega}{\omega_{\mathrm{c}}}\int
\limits^{\infty}_{\omega/\omega_{\mathrm{c}}}K_{5/3}(y)\D y\right] \ .
\label{eq:re}
\end{equation}

\noindent Equation (\ref{eq:re}) is the required expression for
the frequency spectrum of the radiation from an electron moving
along a trajectory which is a circle. For a small argument
$\omega/\omega_{\mathrm{c}} \ll 1$ we may apply an asymptotic
approximation for the modified Bessel's function and get

\begin{displaymath}
\frac{\omega}{\omega_{\mathrm{c}}}
\int\limits^{\infty}_{\omega/\omega_{\mathrm{c}}}K_{5/3}(y)\D y
\simeq 1.33 \left(\frac{\omega}{\omega_{\mathrm{c}}}\right)^{1/3}
\quad {\mathrm{for}} \quad \omega/\omega_{\mathrm{c}} \ll 1 \ .
\end{displaymath}

\begin{figure}[tb]
\begin{center}
\epsfig{file=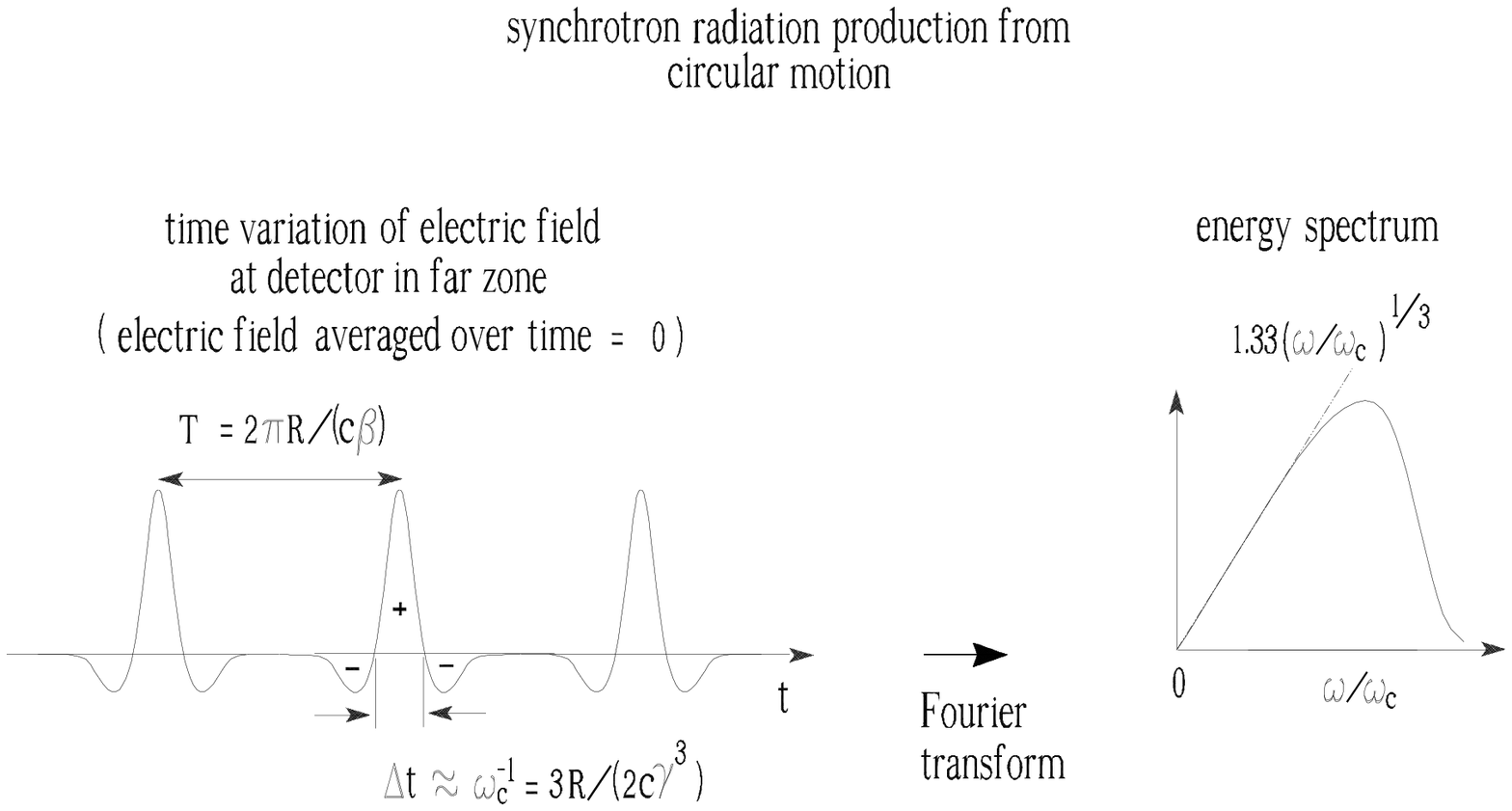,width=0.8\textwidth}
\end{center}
\caption{ Radiation field from an electron moving along a circle
in the time and in the frequency domain} \label{fig:dsm2}
\end{figure}

\begin{figure}[tb]
\begin{center}
\epsfig{file=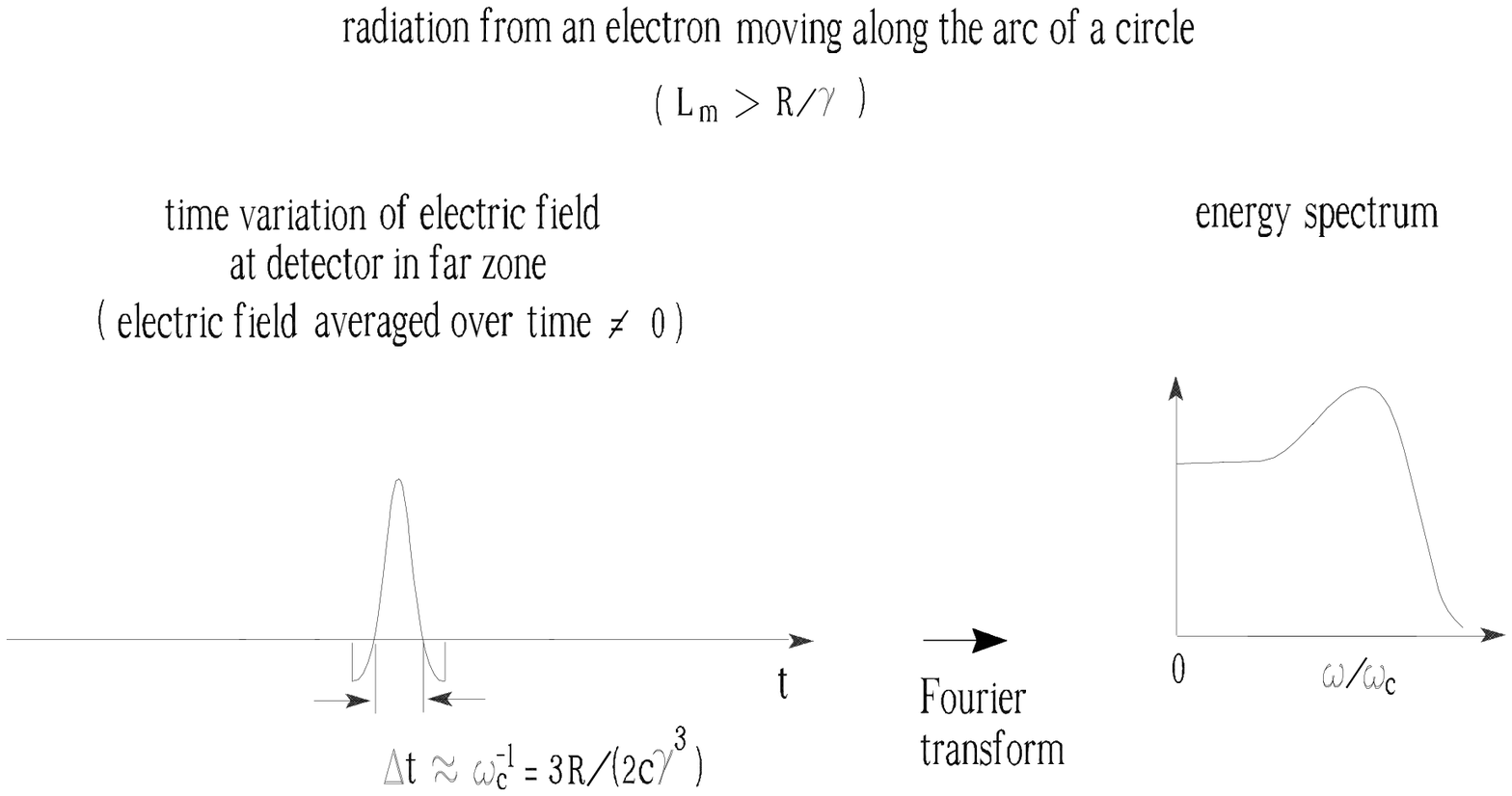,width=0.8\textwidth}
\end{center}
\caption{ Radiation field from an electron moving along an arc of
a circle in the time and in the frequency domain} \label{fig:dsm3}
\end{figure}

\begin{figure}[tb]
\begin{center}
\epsfig{file=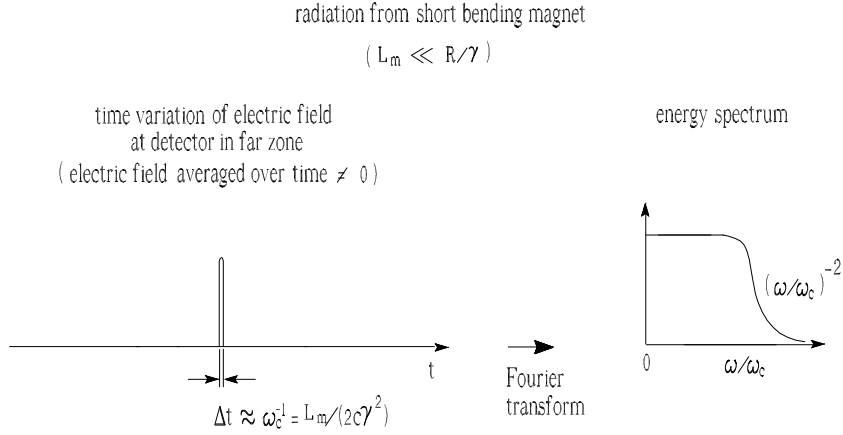,width=0.8\textwidth}
\end{center}
\caption{ Radiation field from an electron moving along a short
bending magnet in the time and in the frequency domain}
\label{fig:dsm4}
\end{figure}

\noindent Most textbooks on synchrotron radiation discuss equation
(\ref{eq:re}). However, no attention is usually paid to the region
of applicability of this equation. Calculations which led to
(\ref{eq:re}) are only valid for an electron moving in a circle.
On the other hand we want to study synchrotron radiation from a
dipole magnet too. Consider what would happen if, instead of an
electron moving in a circle, we had an electron which moves along
the arc of a circle. Fig.~\ref{fig:dsm2}-\ref{fig:dsm4} sketch the
expected frequency spectrum.  These figures provide an
illustration of the transition from the usual synchrotron
radiation by an electron in a circle to radiation by an electron
moving along an arc of a circle. The first sketch
(Fig.~\ref{fig:dsm2}) describes radiation from a circle. It is
useful to consider further the relationship between the time
domain and the frequency domain. The frequency spectrum of the
radiation pulse is given by the expression (\ref{eq:re}): the
Fourier transform at $\omega = 0$, $\bar{E}(0)$, is zero because
$E(t)$ averages to zero (see Fig.~\ref{fig:srap}).

In the case of dipole magnet, things are quite different. The
frequency spectrum may be obtained calculating the Fourier
transform of the time distribution of the electric field at the
observer's position. It can be obtained numerically without much
work by noticing that we can use curves in
Fig.~\ref{fig:sh25a}-\ref{fig:sh05}. As a result, we can define
general properties of the spectrum without any analytical
calculations.  Consider Fig.~\ref{fig:dsm3}, \ref{fig:dsm4}.  As
these curves illustrate, when the electron moves along an arc of a
circle, the spectral distribution does not tend to zero when
$\omega \to 0$. Although this fact may look surprising, this is
quite natural, since the average of the electric field over time
is nonzero when an electron moving along the arc of a circle.

\subsubsection{Radiation from an electron moving along the arc of a
circle}

Many texts on synchrotron radiation theory do not seem to provide
a derivation of the expression for the radiation spectrum from an
electron moving along an arc of a circle. Here we present such a
derivation, which we believe to be quite simple and instructive.
Analytical studies of this problem were first performed in
\cite{arc}. A similar treatment may also be found in the book
\cite{b}. Our consideration is restricted to the simplest case
from an analytical point of view, namely the case of a bending
angle $\phi_{\mathrm{m}}$ very small compared with the $1/\gamma$.
No assumptions on the magnetic field parameters have been made to
derive the radiation spectrum in the form of equation
(\ref{eq:fc}): therefore, our starting point is the same as that
for the radiation spectrum calculation from a circular motion in
the previous subsection. Introducing the notation

\begin{figure}[tb]
\begin{center}
\epsfig{file=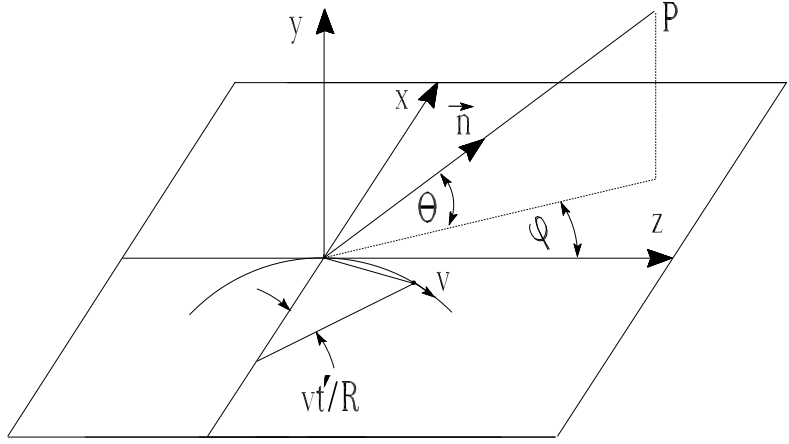,width=0.8\textwidth}
\end{center}
\caption{Geometry for synchrotron radiation from an electron
moving along an arc of a circle } \label{fig:dsm20}
\end{figure}

\begin{displaymath}
\vec{G}(t^{\prime}) = \left[\frac{\vec{n}\times[(\vec{n} -
\vec{\beta})\times\dot{\vec{\beta}}]}{(1 -
\vec{n}\cdot\vec{\beta})^{2}}\right] \ ,
\end{displaymath}

\noindent we rewrite (\ref{eq:fc}) in the form:

\begin{displaymath}
\vec{E}_{\mathrm{r}}(\omega) =
\frac{(-e)}{c\mid\vec{r}_{0}\mid}\int\limits^{\infty}_{-\infty}
\vec{G}(t^{\prime}) \exp[\I\omega(t^{\prime}-
\vec{n}\cdot\vec{R}(t^{\prime})/c)]\D t^{\prime} \ ,
\end{displaymath}

\noindent Now we have to perform the integration. The function
$\vec{G}(t^{\prime})$ is a smooth curve and does not vary very
much across the very narrow bending angle region in the case of a
short magnet: therefore we may replace it by a constant. In other
words, we simply take $\vec{G}(t^{\prime})$ outside the integral
sign and call it $\vec{G}_{0}$. In this case the integral over
$t^{\prime}$ in the latter equation is calculated analytically

\begin{eqnarray}
& \mbox{} & \vec{E}_{\mathrm{r}}(\omega) =
\frac{(-e)\vec{G}_{0}}{c\mid\vec{r}_{0}\mid}\int
\limits^{L_{\mathrm{m}}/(2c\beta)}_{-L_{\mathrm{m}}/(2c\beta)}
\exp[\I\omega(t^{\prime}- \vec{n}\vec{R}(t^{\prime})/c)]\D
t^{\prime}
\nonumber\\
& \mbox{} & =
\frac{(-e)\vec{G}_{0}L_{\mathrm{m}}}{c\mid\vec{r}_{0}\mid}
\frac{\sin\left[(1-\vec{n}\vec{\beta})\omega
L_{\mathrm{m}}/(2c\beta)\right]}{
\left[(1-\vec{n}\cdot\vec{\beta})\omega
L_{\mathrm{m}}/(2c\beta)\right]} \ . \label{eqn:4}
\end{eqnarray}

\noindent Next we compute the quantities required for expressing
explicitly the latter equation. If the system has no azimuthal
symmetry, the angular distribution of radiation depends on two
variables. These may be taken (Fig.~\ref{fig:dsm20}) to be the
angle $\theta$ between $\vec{n}$ and the $x-z$ plane, and the
azimuthal angle $\phi$. We can now write down the components of
the vectors $\vec{\beta}$, $\dot{\vec{\beta}}$, and $\vec{n}$ as:

\begin{displaymath}
\vec{\beta} = \beta\vec{e}_{z} \ , \quad
\dot{\vec{\beta}} =
\dot{\beta}\vec{e}_{x} = \frac{\beta^{2}c}{R}\vec{e}_{x} \ ,
\end{displaymath}

\begin{displaymath}
\vec{n} = \cos\theta\sin\phi\vec{e}_{x} + \sin\theta\vec{e}_{y}
+ \cos\theta\cos\phi\vec{e}_{z} \ .
\end{displaymath}

\noindent These definitions can be used to get the triple vector
product

\begin{eqnarray}
& \mbox{} &
\vec{n}\times[(\vec{n} -
\vec{\beta})\times\dot{\vec{\beta}}] =
\dot{\beta}\left[ \left(\cos^{2}\theta\sin^{2}\phi - 1 +
\beta\cos\theta\cos\phi\right)\vec{e}_{x}
\right.
\nonumber\\
& \mbox{} &
\left.
+ \left(\sin\theta\cos\theta\sin\phi\right)\vec{e}_{y}
+ \left(\cos^{2}\theta\cos\phi\sin\phi -
\beta\cos\theta\cos\phi\right)\vec{e}_{z} \right] \ .
\label{eqn5}
\end{eqnarray}

\noindent We can use these expressions to compute the Fourier
transform (\ref{eqn:4}).  The spectral distribution  is
proportional to the square of electric field and if we follow
through the algebra we find that

\begin{displaymath}
\frac{\D^{2}W}{\D\omega\D\Omega} =
\frac{e^{2}L^{2}_{\mathrm{m}}}{4\pi^{2}R^{2}c}
\frac{(A^{2}_{\sigma} +
A^{2}_{\pi})}{\left(1-\beta\cos\theta\cos\phi\right)^{4}}
\frac{\sin^{2}\left[(1-\beta\cos\theta\cos\phi)\omega
L_{\mathrm{m}}/(2c\beta)\right]}{
\left[(1-\beta\cos\theta\cos\phi)\omega
L_{\mathrm{m}}/(2c\beta)\right]^{2}} \ ,
\end{displaymath}

\noindent where the following notation has been introduced:
$A_{\sigma} = \sin\phi - \beta\sin\theta$ and $A_{\pi} =
\cos\theta\cos\phi$. The quantities $A_{\sigma}$, $A_{\pi}$ are
characteristics of the $\sigma -$ and $\pi -$ components of the
single-particle linearly polarized radiation.

We emphasize the following features of this result. The radiation
is very much collimated in the forward direction, most of the
energy being emitted within the cone $\Delta\Omega =
\Delta\phi\Delta\theta$ of $\gamma^{-2}$. The radiation maximum is
reached at $\theta = 0, \ \phi = 0$. Spectral properties of the
radiation are defined by the function $[\sin^{2}
\omega/\omega_{\mathrm{c}}]/[\omega/\omega_{\mathrm{c}}]^{2}$,
where the critical frequency is $\omega_{\mathrm{c}} =
2c/[(1-\beta\cos\theta\cos\phi)L_{\mathrm{m}}]$. The spectrum from
an electron in a "short" magnet is practically "white noise"
spreading from zero up to the frequencies $\omega \simeq
\omega_{\mathrm{c}}$. As one can see, the spectral properties of
radiation considered here are significantly different from those
of synchrotron radiation from an electron moving in a circle
\footnote{There is an obvious analogy between the synchrotron
radiation from "short" magnet and radiation from bremsstrahlung
effect.  The mathematics of the two problems is essentially the
same}.

\subsection{Limitations of standard results}

As already pointed out, we can  decompose  the CSR spectrum,
$P(\omega)$, into the product of the square modulus of bunch
form-factor $\mid\bar{F}(\omega)\mid^{2}$ with the radiation
spectrum from one electron $p(\omega)$. To find
$\mid\bar{F}(\omega)\mid^{2}$ from $P(\omega)$, one needs the
quantity $p(\omega)$, which is usually very difficult to know with
great accuracy within the long wavelength range. Actually, as
already discussed, the well-known theory of synchrotron radiation
uses approximations which are valid under certain conditions,
inappropriate under others. In particular, Schwinger approach
relies on several assumptions, which we review once again: first,
the observer is located in such a way that the velocity term in
the Lienard-Wiechert formula can be completely neglected and that
the unit vector $\vec{n}$ can be considered constant throughout
the electron evolution. Second, a circular trajectory is
postulated. Finally, no aperture limitations is considered at all.
These assumptions must be analyzed in order to understand how the
CSR pulse is altered in realistic situation. In this subsection we
will deal separately with all of them.

Let us imagine that our electron bunch moves along an arc of a
circle and that there is no aperture limitation. We can take into
account, then, effects as a finite distance between  the source
and observer, a finite magnet length  and the presence of velocity
fields by means of (\ref{eq:csr2}) where the expression for the
electric field is given by the strict Lienard-Wiechert formula.
Analytical methods are of limited use in the study of CSR in the
near zone, and numerical calculations must be selected.  The
application of similarity techniques allows one to present
numerical results in such a way that they are both general and
directly applicable to the calculation of specific device
situations. The use of similarity techniques enables one not only
to reduce the number of parameters but also to reformulate the
problem in terms of variables possessing a clear physical
interpretation. Each physical factor influencing the CSR
production is matched by its own reduced parameter. For the effect
under study this reduced parameter is a measure of the
corresponding physical effect. When some effect becomes less
important for coherent radiation, it simply falls out of the
number of the parameters of the problem.

For our present purposes we would like to concentrate completely
on the temporal structure of the CSR pulse. For any CSR pulse
there exist some characteristic time, which determines the CSR
pulse duration. Its magnitude is of order of the inverse of the
frequency spread in the CSR beam. The behavior of the CSR pulse
profile as a function of dimensionless parameters provides
information on the spectrum distortion. To be specific, we
consider the case of an electron beam with a Gaussian axial
profile of the current density. When comparing the temporal
structure of CSR pulses, it is convenient to use a normalized
field amplitude: here, the normalization is performed with respect
to the maximal field amplitude. The rms electron pulse duration is
assumed to be large $\sigma_{\mathrm{T}} \gg R/(c\gamma^{3})$ and
we focus on the radiation pulse seen about the tangent to the
orbit at the middle point of magnet. When the vertical angle
$\theta = 0$ the normalized coherent field amplitude
$E_{x}(t)/E_{\mathrm{max}}$ is a function of six dimensional
parameters:

\begin{displaymath}
t \ , \quad v \ , \quad R \ , \quad L_{\mathrm{m}} \ , \quad
\sigma_{\mathrm{T}} \ , \quad \mid\vec{r}_{0}\mid \ .
\end{displaymath}

\noindent  It is relevant to note that only two dimensions (length
and time) are sufficient for a full description of the field
profile: after appropriate normalization it is a function of four
dimensionless parameters only:

\begin{equation}
E_{x}(t)/E_{\mathrm{max}} = \hat{E}_{x} = D(\hat{t},\hat{\sigma},
\hat{\rho},\hat{r}_{0}) \ ,
\label{eq:n1}
\end{equation}

\noindent where $\hat{t} = t/\sigma_{\mathrm{T}}$ is the
dimensionless time, $\hat{\sigma} =
\omega_{0}\sigma_{\mathrm{T}}\gamma^{3}$ is the dimensionless
electron pulse duration, $\mid\vec{r}_{0}\mid
(c\sigma_{\mathrm{T}})^{-1/3}R^{-2/3}$ is the dimensionless
distance between source and observer, $\hat{\rho} =
\phi^{3}_{\mathrm{m}}/(6\omega_{0}\sigma_{\mathrm{T}})$ is the
magnet length parameter. In the general case the universal
function $D$ should be calculated numerically by means of strict
Lienard-Wiechert formulas.

The changes of scale performed during the normalization process,
mean that we are measuring time, bunch length, distance from the
source and magnet length as multiples of "natural" CSR units.
There is a physical reason for being able to write the field
profile as in (\ref{eq:n1}): let us explain this fact beginning
with a qualitative analysis of the radiation from an electron
moving in a circle, in the long wavelength asymptotic. Synchrotron
radiation is emitted from a rather small area and we need to
determine this area for observers whose detection systems collect
information over a long time period $\sigma_{\mathrm{T}}$. The
radiation pulse length is equal to the time taken for the electron
to travel along any arc $AB$, reduced by the time taken for the
radiation to travel directly from the arc extreme $A$ to $B$.
Between point $A$ and point $B$ we have a deflection angle $\phi$,
so that $\sigma_{\mathrm{T}} \simeq R\phi/(\beta c) -
2R\sin(\phi/2)/c$, and $\sin(\phi/2)$ can be approximated by
$\sin(\phi/2) \simeq \phi/2 - \phi^{3}/48$ for small angles. Then
the pulse duration reduces to $\sigma_{\mathrm{T}} \simeq
R\phi^{3}/(24c)$. The radiation source extends over some finite
length $R\phi \simeq L_{\mathrm{f}} =
(c\sigma_{\mathrm{T}})^{1/3}R^{2/3}$ along the particle path. We
see that the reduced distance can be expressed as $\hat{r}_{0} =
\mid\vec{r}_{0}\mid/L_{\mathrm{f}}$. One can find that the ratio
$(R\phi_{\mathrm{m}})^{3}/(6L_{\mathrm{f}}^{3})$ is equal to
$\hat{\rho}$, which we use now as a measure of finite magnet
length effects.

\subsubsection{Finite distance effects}

Let us imagine that our electron bunch moves in a real circle and
there is no aperture limitation. We can first consider the
contribution given by the acceleration field alone and then focus
on the contributions by the velocity field. When the electron
bunch radiates from a circle the spectrum  is obviously
independent of the $\hat{\rho}$ parameter. In the long wavelength
asymptotic $(\hat{\sigma} \gg 1)$ the acceleration field is
described by one dimensionless parameter only:  $\hat{r}_{0}$,
where $\hat{r}_{0} =
\mid\vec{r}_{0}\mid(c\sigma_{\mathrm{T}})^{-1/3}R^{-2/3}$ is the
dimensionless distance between radiation source and observer.

The region of applicability of Schwinger's formulas requires the
dimensionless distance to have a large value $\hat{r}_{0} \gg 1$.
In fact, as previously explained, the radiation source extends
over some finite length $R\phi \simeq L_{\mathrm{f}}$, and this
length corresponds to a transverse size of the radiation source $d
\simeq R\phi^{2}$. The vector $\vec{n}$ changes its orientation
between point $A$ and point $B$ of an angle of order $d/r_{0}$,
where $r_{0}$ is the distance between source and observer. Our
estimates show that in the case $d/r_{0} \ll \phi$, the vector
$\vec{n}$ in Lienard-Wiechert formula is almost constant when the
electron moves along the formation length $R\phi$. Thus, we can
conclude  that the unit vector $\vec{n}$ can be considered
constant throughout the electron evolution only if $r_{0} \gg
L_{\mathrm{f}}$.

\begin{figure}[tb]
\begin{center}
\epsfig{file=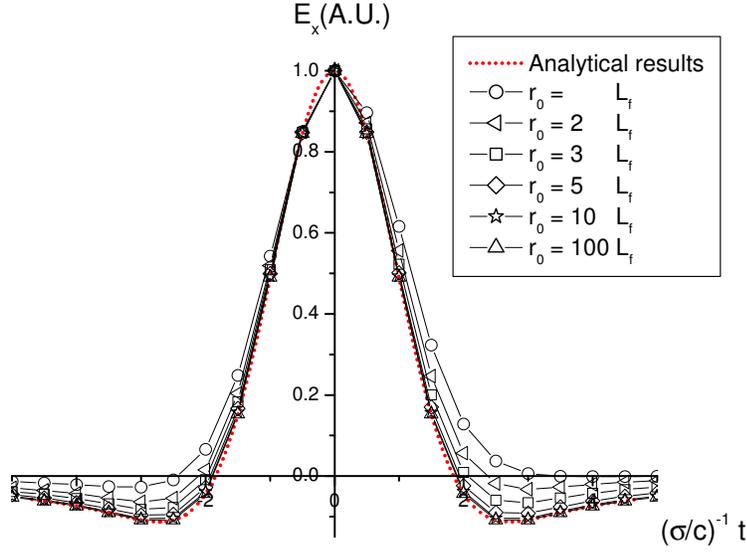,width=0.8\textwidth}
\end{center}
\caption{Time structure of a CSR pulse from a Gaussian electron
bunch moving in a circle at different reduced distance between
source and observer.  Here $\theta = 0$.  The dashed curve is
calculated within the far zone approximation (\ref{eq:csr14}).
Numerical calculations have been performed with strict
Lienard-Wiechert formula } \label{fig:comp}
\end{figure}

\begin{figure}[tb]
\begin{center}
\epsfig{file=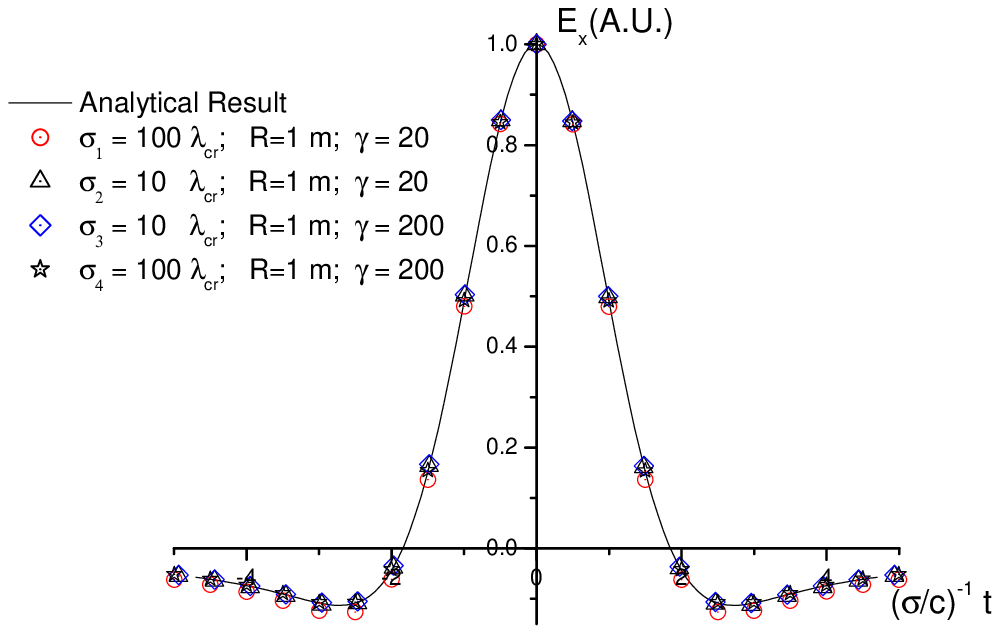,width=0.8\textwidth}
\end{center}
\caption{Illustration of similarity techniques. Time structure of
a CSR pulse from a Gaussian electron bunch moving in a circle for
various sets of  parameters. The reduced distance is held
constant, $\hat{r}_{0} = 15$. The solid curve is calculated within
the far zone approximation (\ref{eq:csr14}). Numerical
calculations have been performed with strict Lienard-Wiechert
formula } \label{fig:final100}
\end{figure}

\begin{figure}[tb]
\begin{center}
\epsfig{file=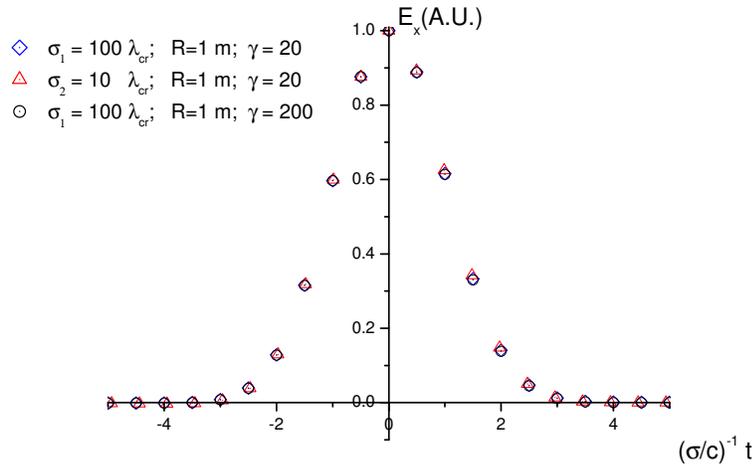,width=0.8\textwidth}
\end{center}
\caption{Illustration of similarity techniques. Time structure of
a CSR pulse from a Gaussian electron bunch moving in a circle for
various sets of parameters. The reduced distance is held constant,
$\hat{r}_{0} = 0.15$.  Numerical calculations have been performed
with strict Lienard-Wiechert formula } \label{fig:final1}
\end{figure}

The results of numerical calculations for several values of
$\hat{r}_{0}$ are presented in Fig. \ref{fig:comp}. Calculations
have been performed using the strict Lienard-Wiechert formula. The
plots in Fig. \ref{fig:comp} give an idea of the region of
validity of the far zone approximation considered above. At large
distance the CSR pulse profile is simply the far zone profile
(\ref{eq:csr14}). One can see that (\ref{eq:csr14}) works well at
$\hat{r}_{0} = 100$. Then, at $\hat{r}_{0} = 3$, the CSR pulse
envelope is visibly modified. As the distance is decreased, the
difference between the approximate and the strict pulse profile
becomes significant. From a practical point of view this set of
plots  covers all the region of interest for the distance between
observer and sources.

To check that no mistakes have been made in our similarity
techniques we evaluate the normalized CSR pulse profile for
several sets of problem parameters. The reduced distance is held
constant. In the present example, these values are $\hat{r}_{0} =
0.15, 15$. The plots are calculated, numerically, using the strict
Lienard-Wiechert formula.  Fig.  \ref{fig:final100}, Fig.
\ref{fig:final1} show such plots.  It is seen that there is a good
agreement with the prediction that the acceleration field from a
circle in long wavelength asymptotic is a function of the reduced
distance only.

\subsubsection{Velocity field effects}

Usual theory of synchrotron radiation is based on the assumption
that acceleration field dominates and all the results presented
above refer to the case when there is no influence of the velocity
field on the detector.  The acceleration field dominates in the
far zone only, and we want to study near zone effects too.  The
physics of coherent effects studied by means of general
expressions for the Lienard-Wiechert fields is much richer than
that of the simplified radiation field model considered above. In
the long wavelength asymptotic the velocity part of the coherent
electric field from a particle in a circle is a function of two
dimensionless parameters:  the reduced distance parameter
$\hat{r}_{0}$ and the reduced electron pulse duration parameter
$\hat{\sigma}$:

\begin{equation}
E_{\mathrm{(v)}}(t)/E_{\mathrm{max}} = \hat{E}_{\mathrm{(v)}} =
D(\hat{t},\hat{\sigma}, \hat{r}_{0}) \ ,\label{eq:n2}
\end{equation}

\noindent where the normalization of the velocity field is
performed with respect to the maximal acceleration field
amplitude.

\begin{figure}[tb]
\begin{center}
\epsfig{file=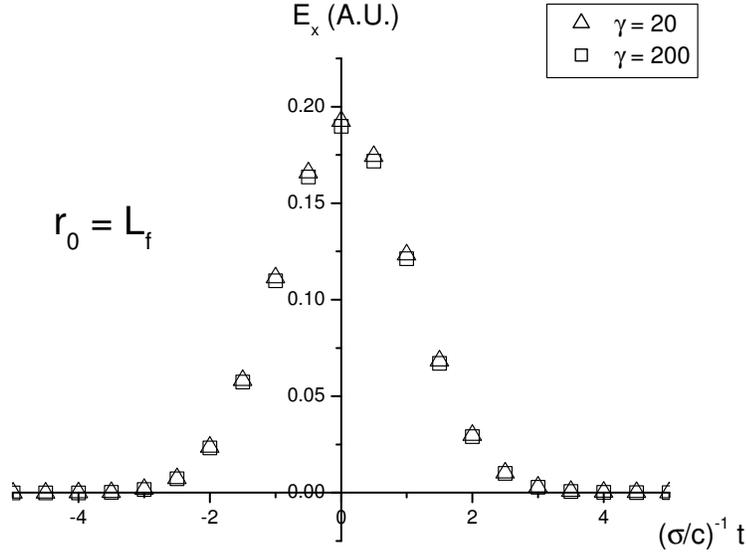,width=0.8\textwidth}
\end{center}
\caption{Illustration of similarity techniques. Electric field
pulse due to velocity term from a Gaussian electron bunch moving
in a circle for various sets of parameters. The reduced distance
and reduced bunch length are held constant, $\hat{r}_{0} = 1$ and
$\hat{\sigma}_{\mathrm{T}} = 100$ respectively. Numerical
calculations have been performed with strict Lienard-Wiechert
formula } \label{fig:uncmp}
\end{figure}

To show that this is correct we can perform a simple check. In
Fig. \ref{fig:uncmp} we present numerical calculation results for
the velocity field for different sets of parameters. The reduced
distance and the reduced electron pulse duration are held
constant. It is seen that all numerical results agree rather well.
Therefore the plots in Fig.  \ref{fig:uncmp} convince us that the
dimensionless equation (\ref{eq:n2}) provides an adequate
description of the numerical calculations.

\begin{figure}[tb]
\begin{center}
\epsfig{file=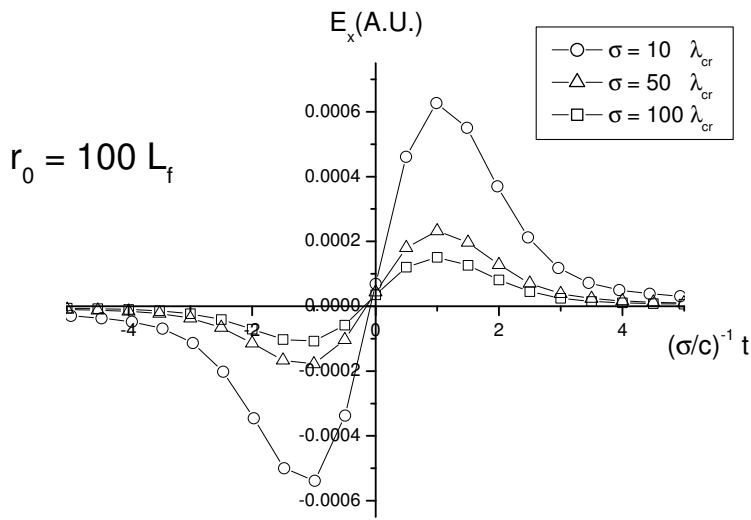,width=0.8\textwidth}
\end{center}
\caption{Electric field pulse due to velocity term from a Gaussian
electron bunch moving in a circle at different reduced bunch
length. The reduced distance is held constant, $\hat{r}_{0} =
100$. Here $\theta = 0$ and the normalization is performed with
respect to the maximal acceleration field amplitude }
\label{fig:th100}
\end{figure}

\begin{figure}[tb]
\begin{center}
\epsfig{file=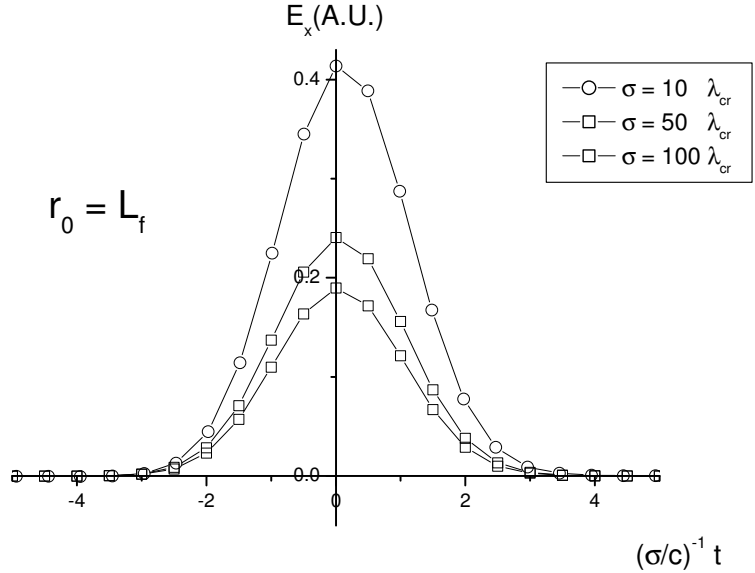,width=0.8\textwidth}
\end{center}
\caption{Electric field pulse due to velocity term from a Gaussian
electron bunch moving in a circle at different reduced bunch
length.  The reduced distance is held constant,  $\hat{r}_{0} =
1$.  Here $\theta = 0$ and the normalization is performed with
respect to the maximal acceleration field amplitude }
\label{fig:th13}
\end{figure}

Fig. \ref{fig:th100} and Fig. \ref{fig:th13} show the dependence
of the normalized velocity field amplitude on the value of the
reduced electron pulse duration for different values of the
reduced distance. Using these plots we can give a quantitative
description of the region of applicability of the radiation field
model. It is seen from the plots in Fig. \ref{fig:th13} that in
the near zone we cannot neglect the influence of the velocity
field on the detector.

Let us now estimate the importance of the velocity field effect.
Let us start considering the total velocity field pulse as a
superposition of single particle fields at a given position in the
far zone. To calculate the integral one should take into account
the property of its kernel. In the far zone the velocity field
from one electron is close to an antisymmetric function and the
average of the electric field over time is close to zero. The
approach used in section 3.1 can be also used  to compute the
coherent velocity field. Under the long electron bunch condition
the kernel (velocity field from one electron) can be substituted
by its asymptotic behavior. If we wish to estimate the normalized
amplitude of the coherent velocity field  we can get it simply by
dividing the asymptotic behavior of the velocity field kernel by
the asymptotic behavior of the acceleration field kernel, so that,
in the far zone, the normalized velocity field is of order
$E_{\mathrm{v}}/E_{\mathrm{acc}} \simeq
R/(\gamma^{2}\theta^{2}r_{0})$, where $\theta \simeq
(c\sigma_{\mathrm{T}}/R)^{1/3}$ is the natural synchrotron
radiation opening angle with frequency $\omega \simeq
\sigma^{-1}_{\mathrm{T}} \ll c\gamma^{3}/R$. Using normalized
variables we get

\begin{displaymath}
\hat{E}_{\mathrm{(v)}}  \simeq
\hat{r}^{-1}_{0}\hat{\sigma}^{-2/3} \quad {\mathrm{for}}
\quad \hat{\sigma}, \ \hat{r}_{0} \gg 1 \ .
\end{displaymath}

\noindent As we can see from Fig. \ref{fig:th100}, numerical
calculations in the far zone confirm this simple physical
consideration. The value of $\hat{E}_{\mathrm{v}}$ that we can
expect is found remembering that, in the example given in Fig.
\ref{fig:th100}, $\hat{r}_{0} = 100, \ \hat{\sigma} = 100$;
therefore the normalized velocity field would be about 0.0004,
which is the same order of magnitude as results of numerical
calculations (0.0002). Also, $\hat{E}_{\mathrm{v}}$ varies roughly
as $\hat{\sigma}^{-2/3}$.

The normalized velocity field amplitude decreases, as we see from
our estimations, linearly with distance, which means that if we
are in the near zone at $\hat{r}_{0} = 1$ there will be
$\hat{E}_{\mathrm{v}} \simeq 0.04$. Numerical calculation results
show that, at the value $\hat{r}_{0} = 1$, we have
$\hat{E}_{\mathrm{v}} = 0.2$. We should say that our approximate
treatment of coherent velocity field breaks down once source and
observer get as close as they are at $\hat{r}_{0} \simeq 1$. We
have, however, a simple explanation for that. In addition to the
antisymmetric part of kernel we have just described, there is also
a symmetric part.  When source and observer are far apart the
observer sees only the antisymmetric part and the average of
electric field from one electron is close to zero. At very close
distances there begins to be some extra symmetric part of the
kernel. This symmetric field, which also varies with the
separation, should be, of course, included in more precise
estimates.

\subsubsection{Finite magnet length effects}

\begin{figure}[tb]
\begin{center}
\epsfig{file=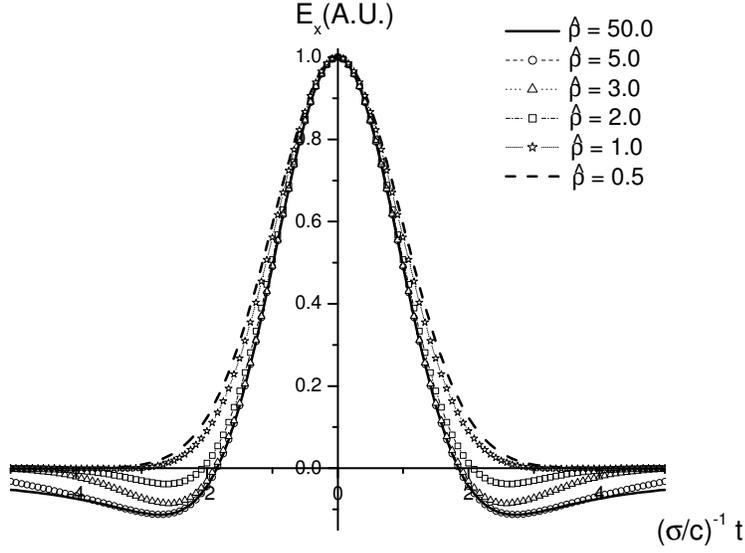,width=0.8\textwidth}
\end{center}
\caption{Time structure of a CSR pulse in the far zone from a
Gaussian electron bunch moving along an arc of a circle at
different reduced magnet lengths.  Here $\theta = 0$. Calculations
have been performed with anaytical formula (\ref{eq:csr95}).
Results at $\hat{\rho} = 50$ (solid curve) agree with our
analytical expression (\ref{eq:csr14}) for circular motion within
1\% } \label{fig:tcom}
\end{figure}

Until now we have considered the case in which electrons moves in
a circle. Here we will generalize this situation and study the
case where the electrons move along an arc of a circle. In order
to fully characterize the magnet length required to assure the
accuracy of the circular motion model, it is necessary to specify
the distortion of Schwinger spectrum  associated with all
interesting magnet lengths in an experiment. Normalizing equation
(\ref{eq:csr95}), we obtain that, under the far field
approximation, the  CSR pulse profile is a function of only one
dimensionless parameter $\hat{\rho} =
\phi_{\mathrm{m}}^{3}/(6\omega_{0}\sigma_{\mathrm{T}})$, where
$\phi_{\mathrm{m}}$ is the magnet angular extension. The region of
applicability of  Schwinger's formulas require the parameter
$\hat{\rho}$ to have a large value $\hat{\rho} \gg 1$. Using the
plots presented in Fig. \ref{fig:tcom}, one can characterize
quantitatively the region of applicability of the circular motion
model.

\subsubsection{Diffraction effects}

Another problem concerns radiation spectrum distortions: this has
to do with aperture limitations. In real accelerators, the long
wavelength synchrotron radiation from bending magnets integrates
over many different vacuum chamber pieces with widely varying
aperture. The  SR spectrum formed by such a system will be
perturbed by the presence of an aperture and we should seek a
mathematical mean for predicting these perturbations.

\begin{figure}[tb]
\begin{center}
\epsfig{file=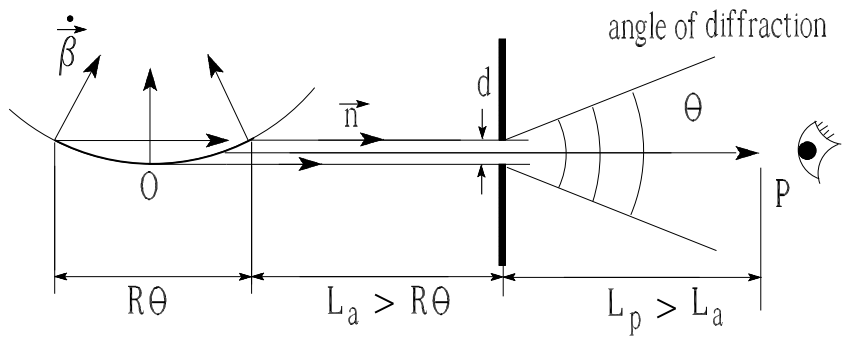,width=0.8\textwidth}
\end{center}
\caption{ Depiction of the effects of an aperture limitation. CSR
from a circle allows us to measure the bunch profile form-factor
in a frequency range down to $\omega_{\mathrm{a}}$ only}
\label{fig:dsm17}
\end{figure}

One may wonder how the existence of an aperture in an accelerator
vacuum chamber and in a beam line, which is relatively far from
the source, can influence the synchrotron radiation spectrum. To
illustrate this point, we consider the situation shown in Fig.
\ref{fig:dsm17}. The object of interest is an electron moving in a
circle. Between the observer and the source there is an aperture
with a characteristic dimension $d$. Qualitatively, an observer
looking at a single electron is presented with a cone of radiation
characterized by an aperture angle of order $\theta \simeq
\sqrt{d/R}$. Fig. \ref{fig:dsm17} shows part of the trajectory of
an electron travelling along an arc of a circle of radius $R$. At
first glance this intuitive argument indicates that the frequency
spectrum may be obtained by reducing the situation shown in Fig.
\ref{fig:dsm17} to Fig. \ref{fig:dsm11} and simply calculating
radiation from an electron moving along a trajectory which is an
arc of a circle. However, the situation is more complicated than
that. The field at $P$ can be represented as a sum of two parts -
the field due to the source plus the field due to the wall, i.e.
due to the motions of the charges in the wall. The latter kind of
radiation field can be described in terms of diffraction theory.
Diffraction is the process by which radiation is redirected near
sharp edges. As it propagates away from these sharp edges, it
interferes with nearby non-diffracted radiation, producing
interference patterns. A finite aperture introduces, therefore,
diffraction effects specific to the geometry and clearly dependent
on the wavelength. For structures such as pinholes it is found
that these diffraction patterns propagate away from the structure
at angles of order $\theta_{\mathrm{d}} \simeq c/\omega d$, where
$d$ is the characteristic aperture dimension.  The region of
applicability for the far diffraction zone where effects play a
significant role is given by the relation $L_{\mathrm{a}} \gg
L_{\mathrm{d}} \simeq \omega d^{2}/c$, where $L_{\mathrm{a}}$ is
the distance between observer and aperture. When the wavelength is
about $R\theta^{3} \simeq (d^{3}/R)^{1/2}$, the latter condition
transforms to $L_{\mathrm{a}} \gg L_{\mathrm{d}} \simeq R\theta$.
This simple physical argument indicates that the situation shown
in Fig. \ref{fig:dsm17} at $L_{\mathrm{a}} > R\theta \simeq
(Rd)^{1/2}$ cannot be reduced to the situation shown in Fig.
\ref{fig:dsm11}. The significance of the discussed limitation
cannot be fully appreciated until we determine typical values of
the parameters that can be expected in practice. For example, if
$R = 3 {\mathrm{m}}$ and $\lambda = 100 \mu{\mathrm{m}}$,  $\theta
\simeq 0.03$, so that at distance greater than 10 cm from the
aperture, a 1 cm diameter hole would be sufficiently narrow to
perturb significantly the synchrotron radiation spectrum at that
wavelength.

The computation of spectrum distortions for practical accelerator
applications is a rather difficult problem. One encounters vacuum
chamber components which require special attention and
sophisticated techniques should be developed to deal with this
situation. The character of spectrum distortions largely depends
on the actual geometry and on the material of the vacuum chamber,
so that we may expect significant complications in the
determination of these distortions along a vacuum enclosure of the
actual device. The vacuum chamber of an accelerator presents a too
complicated geometry to allow an analytical expression for
spectrum distortions. In principle each section of the vacuum
chamber must be treated individually. By employing two or
three-dimensional numerical codes it may be possible to determine
the spectrum for a particular component. Yet, every accelerator is
somewhat different from another and will have its own particular
spectrum distortions. For this reason, in this discussions we
focused only on basic ideas about aperture limitations.

\section{Radiation from an undulator}

To optimally meet the needs for beam diagnostic measurements with
the features of synchrotron radiation, it is desirable to provide
specific radiation characteristics that cannot be obtained from
bending magnets but require special magnetic systems. To generate
specific synchrotron radiation characteristics, radiation is often
produced by special insertion devices installed along the particle
beam path.  An insertion device does not introduce a net
deflection of the beam and we may therefore choose any arbitrary
field strength which is technically feasible to adjust the
radiation spectrum to experimental needs. In this Section we
discuss the basic theory of one such devices, a planar undulator,
and its applications to CSR diagnostics.

\subsection{Basic theory of undulator radiation}

\begin{figure}[tb]
\begin{center}
\epsfig{file=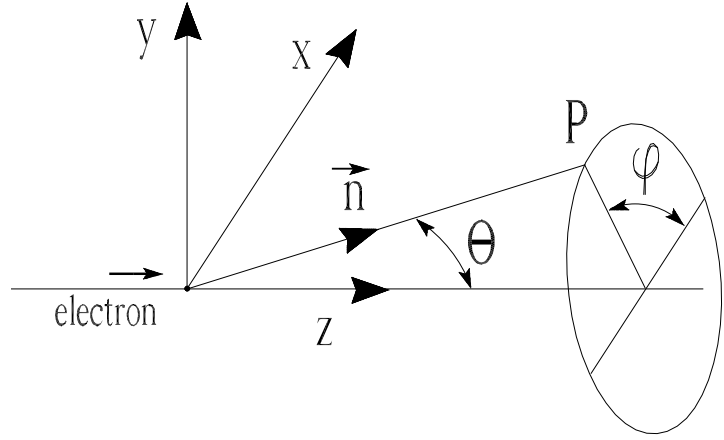,width=0.8\textwidth}
\end{center}
\caption{Radiation geometry for the undulator field}
\label{fig:dsm21}
\end{figure}

The magnetic field on the axis of a planar undulator is given by

\begin{displaymath}
\vec{H}(z) =  \vec{e}_{y}H_{\mathrm{w}}\sin(k_{\mathrm{w}}z) \ ,
\end{displaymath}

\noindent where $\vec{e}_{y}$ is the unit vector directed along
the $y$ axis of the Cartesian coordinate system $(x,y,z)$ in Fig.
\ref{fig:dsm21}. The Lorentz force is used to derive the equation
of motion of an electron with energy $\gamma m_{\mathrm{e}}c^{2}$
in the presence of a magnetic field. Integration of this equation
gives

\begin{displaymath}
\vec{v}_{x}(z) = \vec{e}_{x}c
\theta_{\mathrm{w}}\cos(k_{\mathrm{w}}z) \ ,
\end{displaymath}

\noindent where $\theta_{\mathrm{w}} = K/\gamma$ and
$K = eH_{\mathrm{w}}/(m_{\mathrm{e}}c^{2}k_{\mathrm{w}})$
is the undulator parameter. It is useful to present another
form of this expression convenient for numerical calculations:
$K = 0.093\times H_{\mathrm{w}}({\mathrm{kGs}})\times
\lambda_{\mathrm{w}}({\mathrm{cm}})$ where $\lambda_{\mathrm{w}} =
2\pi/k_{\mathrm{w}}$. The transverse and longitudinal velocity
components are related by
$(\dot{z})^{2} = \beta^{2}c^{2} - (\dot{x})^{2}$, which gives

\begin{displaymath}
\beta_{z} = \beta_{\mathrm{av}} - \frac{K^{2}}{4\gamma^{2}}\cos
2k_{\mathrm{w}}z \ ,
\end{displaymath}

\noindent where $\beta_{\mathrm{av}} = \beta[1 -
K^{2}/(4\gamma^{2})]$ is the mean velocity of the electron along
the undulator.

\begin{figure}[tb]
\begin{center}
\epsfig{file=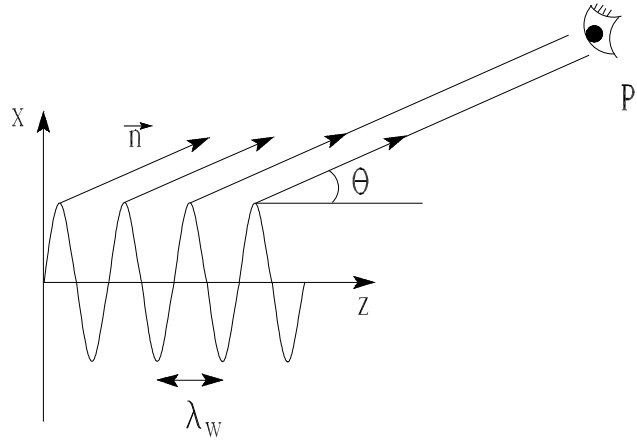,width=0.8\textwidth}
\end{center}
\caption{Synchrotron radiation production from successive poles}
\label{fig:dsm18}
\end{figure}

It is generally assumed that the distance between the undulator
and the observation point is much larger than the undulator
length. This assumption is usually referred to as  far field
approximation. Suppose this condition is met. Denoting with
$\vec{r}$, as usual, the vector from the electron position at the
retarded time to the observation point and with $\vec{r}_{0}$ the
vector from the coordinate origin, placed somewhere inside the
undulator, to the observation point we compute $\vec{r} \simeq
\vec{r}_{0} - \vec{n}\cdot\vec{R}$ with $\vec{n} =
\vec{r}_{0}/\mid\vec{r}_{0}\mid$. The unit vector from the
retarded source particle to the observer is, from Fig.
\ref{fig:dsm21}

\begin{displaymath}
\vec{n} = \cos\phi\sin\theta\vec{e}_{x}
+ \sin\phi\sin\theta\vec{e}_{y}
+ \cos\theta\vec{e}_{z} \ .
\end{displaymath}

\noindent The vector $\vec{R}$, which describes the motion of the
electron in the $x-z$ plane, can be obtained from the previous
treatment:

\begin{displaymath}
x(t) = \frac{K}{k_{\mathrm{w}}\gamma}\sin
(k_{\mathrm{w}}\beta_{\mathrm{av}}ct)\ , \quad
z(t) = \beta_{\mathrm{av}}ct -
\frac{K^{2}}{8k_{\mathrm{w}}\gamma^{2}}\sin
(2k_{\mathrm{w}}\beta_{\mathrm{av}}ct) \ .
\end{displaymath}

\noindent Let us find the Fourier harmonics of the radiation field
from an electron moving along the undulator. As in section 3.2,
our starting point is equation (\ref{eq:fc}) since, once again,
this is a rigorous equation under the assumption of a large
distance between observer and radiating electron. Note that the
integration in (\ref{eq:fc}) is to be performed over all time but
in practice, without loss of generality, it can be restricted to
the time during which the electron is passing through the
undulator.  We can write:

\begin{equation}
\vec{E}_{\mathrm{r}}(\omega) =
\frac{(-e)}{c\mid\vec{r}_{0}\mid}\int\limits^{N_{\mathrm{w}}T}_{0}
\left[\frac{\vec{n}\times[(\vec{n} -
\vec{\beta})\times\dot{\vec{\beta}}]}{(1 -
\vec{n}\cdot\vec{\beta})^{2}}\right] \exp[\I\omega(t^{\prime}-
\vec{n}\cdot\vec{R}(t^{\prime})/c)]\D t^{\prime} \ ,
\label{eq:fcu}
\end{equation}

\noindent  where $T =
\lambda_{\mathrm{w}}/(c\beta_{\mathrm{av}})$, and $N_{\mathrm{w}}$
is the number of undulator periods. The integral in the latter
expression can be represented as a sum of $N_{\mathrm{w}}$
integrals over each undulator period (see Fig. \ref{fig:dsm18}).
Integrating by parts  (\ref{eq:fcu})  and taking into account the
sum of the geometric progression

\begin{displaymath}
\sum^{N_{\mathrm{w}}}_{n=1}\exp(\I n\phi) = \frac{\sin
N_{\mathrm{w}}\phi/2}{\sin \phi/2}\exp\I(N_{\mathrm{w}}+1)\phi/2
\end{displaymath}

\noindent we find

\begin{eqnarray}
& \mbox{} & \vec{E}_{\mathrm{r}}(\omega) =
\frac{(-e)}{c\mid\vec{r}_{0}\mid}\left\{ \left.
\frac{\vec{n}\times[\vec{n}\times\vec{\beta}]}{(1 -
\vec{n}\cdot\vec{\beta})} \exp\left[\I\omega(t^{\prime}-
\vec{n}\cdot\vec{R}(t^{\prime})/c)\right]
\right|^{t^{\prime}=N_{\mathrm{w}}T}_{t^{\prime}=0} \right.
\nonumber\\
& \mbox{} &
\left.
-
\I\omega
\frac{\sin
N_{\mathrm{w}}\omega(1-\beta_{\mathrm{av}}n_{z}))T/2}{\sin
\omega(1-\beta_{\mathrm{av}}n_{z})T/2}
\exp\left[\I(N_{\mathrm{w}}-1)
\omega(1-\beta_{\mathrm{av}}n_{z})T/2\right]
\right.
\nonumber\\
& \mbox{} &
\left.
\times
\int\limits^{T}_{0}
\vec{n}\times[\vec{n}\times\vec{\beta}]
\exp\left[\I\omega(t^{\prime}-
\vec{n}\cdot\vec{R}(t^{\prime})/c)\right] \D t^{\prime}\right\} \ .
\label{eq:byp}
\end{eqnarray}

\noindent For our purposes, it is most convenient to express the
latter equation in the form

\begin{eqnarray}
& \mbox{} & \vec{E}_{\mathrm{r}}(\omega) = \frac{2\I
(-e)}{c\mid\vec{r}_{0}\mid(1-\beta_{\mathrm{av}}\cos\theta)}
\exp\left[\I(N_{\mathrm{w}}-1)\omega(1-\beta_{\mathrm{av}}
\cos\theta)T/2\right]
\nonumber\\
& \mbox{} &
\times
\left\{
\vec{n}\times[\vec{n}\times\vec{\beta}_{0}]\frac{
(1 - \beta_{\mathrm{av}}\cos\theta)}
{(1 - \vec{n}\cdot\vec{\beta}_{0})}
\right.
\nonumber\\
& \mbox{} &
\left.
\times
\exp\left[\I\omega(1-\beta_{\mathrm{av}}\cos\theta)T/2\right] \sin\left[
N_{\mathrm{w}}\omega(1-\beta_{\mathrm{av}}\cos\theta)T/2\right]
\right.
\nonumber\\
& \mbox{} &
\left.
-
\left[\omega(1-\beta_{\mathrm{av}}\cos\theta)T/2\right]
\frac{\sin
N_{\mathrm{w}}\omega(1-\beta_{\mathrm{av}}\cos\theta)T/2}{\sin
\omega(1-\beta_{\mathrm{av}}\cos\theta)T/2}
\right.
\nonumber\\
& \mbox{} &
\left.
\times
\frac{1}{T}\int\limits^{T}_{0}
\vec{n}\times[\vec{n}\times\vec{\beta}]
\exp\left[\I\omega(t^{\prime}-
\vec{n}\cdot\vec{R}(t^{\prime})/c)\right] \D t^{\prime}\right\} \ .
\label{eq:byp1}
\end{eqnarray}

\noindent Here $c\vec{\beta}_{0} = c\vec{\beta}(0) =
c\vec{\beta}(N_{\mathrm{w}}T)$ is the initial and final velocity
of the electron. From the shape of the interference function
$\sin^{2}(N_{\mathrm{w}}x)/\sin^{2}(x)$ we predict that important
contributions to the second term will occur when $x =
\omega(1-\beta_{\mathrm{av}}\cos\theta)T/2$ does not differ
significantly from the resonance value $x = \pi$. The maximum of
the absolute value at resonance is, then, proportional to the
number of periods $N_{\mathrm{w}}$ and, on the assumption of
$N_{\mathrm{w}} \gg 1$, we can neglect the first term.

The spectral distribution of the radiation energy is given by
equation (\ref{eq:es}). Taking the square modulus of
(\ref{eq:byp1}) and keeping the resonance term only, we find

\begin{eqnarray}
& \mbox{} &
\frac{\D^{2}W}{\D\omega\D\Omega} =
\frac{e^{2}\omega^{2}}{4\pi^{2}c}
\left[\frac{\sin
N_{\mathrm{w}}\omega(1-\beta_{\mathrm{av}}\cos\theta)T/2}{\sin
\omega(1-\beta_{\mathrm{av}}\cos\theta)T/2}\right]^{2}
\nonumber\\
& \mbox{} &
\times
\left|\int\limits^{T}_{0}
\vec{n}\times[\vec{n}\times\vec{\beta}]
\exp\left[\I\omega(t^{\prime}-
\vec{n}\cdot\vec{R}(t^{\prime})/c)\right] \D
t^{\prime}\right|^{2} \ .
\label{eqn:u1}
\end{eqnarray}

\noindent Let us now analyze the interference function in
(\ref{eqn:u1})

\begin{displaymath}
f(\omega) = \left[\frac{\sin
N_{\mathrm{w}}\omega(1-\beta_{\mathrm{av}}\cos\theta)T/2}{\sin
\omega(1-\beta_{\mathrm{av}}\cos\theta)T/2}\right]^{2}
\end{displaymath}

\noindent and study some of its
consequences.  The whole curve is localized near

\begin{displaymath}
\omega = \omega_{\mathrm{res}} = 2ck_{\mathrm{w}}\gamma^2\left[1 +
\frac{K_{\mathrm{w}}^{2}} {2} + \gamma^2\theta^{2}\right]^{-1} \ .
\end{displaymath}

\begin{figure}[tb]
\begin{center}
\epsfig{file=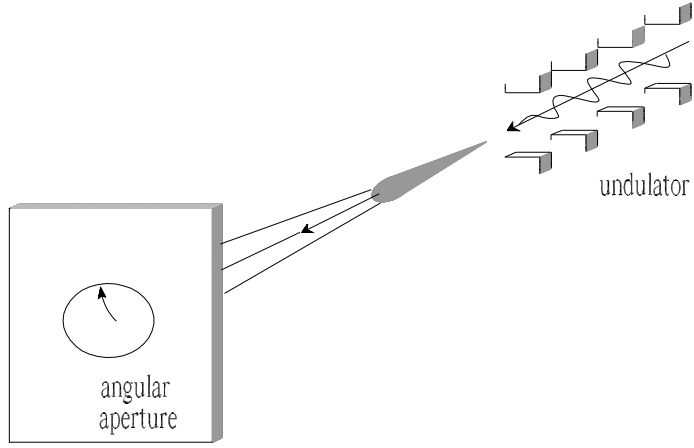,width=0.8\textwidth}
\end{center}
\caption{Undulator radiation with an angular filter}
\label{fig:dsm12}
\end{figure}

\begin{figure}[tb]
\begin{center}
\epsfig{file=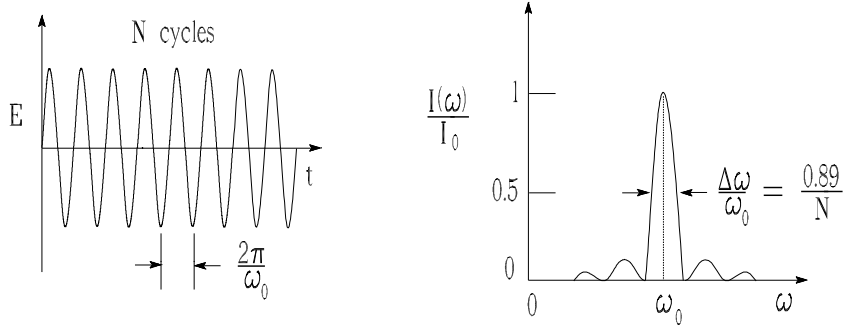,width=0.8\textwidth}
\end{center}
\caption{Radiated wavetrain from a single electron traversing an
undulator, as detected into the central cone, and  corresponding
spectral distribution function} \label{fig:dsm14}
\end{figure}

\noindent The latter condition is called resonance condition and
tells us the radiation frequency as a function of the undulator
period $\lambda_{\mathrm{w}}$, the undulator parameter
$K_{\mathrm{w}}$, the electron energy $\gamma m_{\mathrm{e}}c^2$,
and the polar angle of observation $\theta$. Let us concentrate on
the radiation produced in the forward direction.  In this case
$\theta = 0$ and there is no vertically polarized radiation. By
means of an angular filter, whose principle is shown in Fig.
\ref{fig:dsm12},  we can obtain the undulator radiation from the
central cone. Note that for the radiation within the cone of half
angle

\begin{displaymath}
\theta_{\mathrm{cen}} = \frac{\sqrt{1+K^{2}_{\mathrm{w}}/2}}
{\gamma\sqrt{N_{\mathrm{w}}}} \ ,
\end{displaymath}

\noindent the relative spectral FWHM bandwidth is
$\Delta\omega/\omega = 0.89/N_{\mathrm{w}}$. This means that we
have a very sharp central maximum with very weak subsidiary maxima
on the sides. In fact, it is possible to prove that the intensity
at the next maximum is less than 5\% of the first maximum
intensity (see Fig. \ref{fig:dsm14}). There is another important
feature of the interference function: if the frequency is
increased by any multiple of $\omega_{\mathrm{res}}$, we get other
strong maxima of interference function at $\omega =
2\omega_{\mathrm{res}}, \ 3\omega_{\mathrm{res}}, \
4\omega_{\mathrm{res}}$, and so forth. Near each of these maxima
the pattern of Fig. \ref{fig:dsm14} is repeated.

There is no forward radiation for even harmonics of the
fundamental frequency $\omega_{\mathrm{res}}$. Let us briefly
justify the latter statement. First let us calculate the integral
in (\ref{eqn:u1}). At $\theta = 0$ the triple vector product is

\begin{equation}
\vec{n}\times[\vec{n}\times\vec{\beta}] =
\vec{\beta} - \vec{n}(\vec{n}\cdot\vec{\beta}) =
\beta_{x}\vec{e}_{x} = \frac{K}{\gamma}
\cos(k_{\mathrm{w}}z)\vec{e}_{x} \ .
\label{eq:t1}
\end{equation}

\noindent The exponential factor in (\ref{eqn:u1}) can be evaluated
in a similar way

\begin{equation}
\exp\left[\I m\omega_{0}(t^{\prime}-
\vec{n}\cdot\vec{R}(t^{\prime})/c)\right] =
\exp\left[\I mk_{\mathrm{w}}z + \I Q\sin 2k_{\mathrm{w}}z\right] \ ,
\label{eq:e1}
\end{equation}

\noindent where $m$ is integer,

\begin{displaymath}
Q = K^{2}m\omega_{0}/(8k_{\mathrm{w}}\gamma^{2}) =
m K^{2}_{\mathrm{w}}/(4 + 2K^{2}_{\mathrm{w}}) \ ,
\end{displaymath}

\noindent and $\omega_{0} =
2\gamma^2k_{\mathrm{w}}/[c(1+K^{2}_{\mathrm{w}}/2)]$ is the
resonance frequency of the fundamental harmonic at $\theta = 0$.
Using (\ref{eq:t1}) and (\ref{eq:e1}), together with the expansion

\begin{displaymath}
\exp[\I Q\sin(2k_{\mathrm{w}}z)]
= \sum^{n=+\infty}_{n=-\infty}J_{n}(Q)\exp(2\I nk_{\mathrm{w}}z) \ ,
\end{displaymath}

\noindent where $J_{n}$ is the Bessel function of order $n$, we
obtain:

\begin{eqnarray}
& \mbox{} &
\left|\int\limits^{T}_{0}
\vec{n}\times[\vec{n}\times\vec{\beta}]
\exp\left[\I m\omega_{0}(t^{\prime}-
\vec{n}\cdot\vec{R}(t^{\prime})/c)\right] \D
t^{\prime}\right|^{2} =
\nonumber\\
& \mbox{} &
\frac{K^{2}}{\gamma^{2}}\left|
\sum^{n=+\infty}_{n=-\infty}J_{n}(Q)
\int\limits^{\lambda_{\mathrm{w}}}_{0}
\cos(k_{\mathrm{w}}z)\exp[\I(2n+m)k_{\mathrm{w}}z]\D z\right|^{2} \ ,
\label{eqn:u2}
\end{eqnarray}

\noindent which is non-zero only when $m + 2n \pm 1 = 0$. When $m$
is odd  the spectral and angular density of the radiation energy
emitted at zero angle by a single electron during the undulator
pass is given by the expression

\begin{displaymath}
\left.\frac{\D^{2}W(\omega_{m})}{\D\omega\D\Omega}\right|_{\theta=0}
= \frac{e^{2}N^{2}_{\mathrm{w}}\gamma^{2}m^{2}
A_{\mathrm{JJ}}^{2}K^{2}_{\mathrm{w}}}
{c(1+K^{2}_{\mathrm{w}}/2)^{2}} \frac{\sin^{2}[\pi
N_{\mathrm{w}}(\omega-\omega_{m})/\omega_{m}]} {[\pi
N_{\mathrm{w}}(\omega-\omega_{m})/\omega_{m}]^{2}} \ ,
\end{displaymath}

\noindent where the following notation has been introduced:
$\omega_{m} = m\omega_{0}$,

\begin{displaymath}
A_{\mathrm{JJ}} = [J_{(m-1)/2}(Q)-J_{(m+1)/2}(Q)] \ .
\end{displaymath}

\noindent When deriving the latter expression, we used the
following relation for the Bessel functions: $J_{-n}(Q) =
(-1)^{n}J_{n}(Q)$. Now we can proceed to calculate the energy
radiated into the central cone at the fundamental harmonic.  In
the small angle approximation the solid angle is equal to
$\D\Omega = \theta\D\theta\D\phi$.  Integration of spectral and
angular density over $\omega$ and $\phi$ gives us factors
$\omega_{0}/N_{\mathrm{w}}$ and $2\pi$ respectively.  We also have
integrate over $\theta$ from $0$ to $\theta_{\mathrm{cen}}$.
Altogether, the energy radiated into the central cone by a single
electron is given by

\begin{equation}
\Delta W_{\mathrm{cen}} \simeq \frac{\pi
e^2A_{\mathrm{JJ}}^{2}\omega_{0}K^{2}_{\mathrm{w}}}
{c(1+ K^{2}_{\mathrm{w}}/2)} \ ,
\label{eq:cr}
\end{equation}

\noindent where, in this case, $A_{\mathrm{JJ}} =
[J_{0}(Q)-J_{1}(Q)]$ and $Q = K^{2}_{\mathrm{w}}/(4 +
2K^{2}_{\mathrm{w}})$.

\subsection{Bunch length diagnostic technique based on coherent
radiation from an undulator}

\begin{figure}[tb]
\begin{center}
\epsfig{file=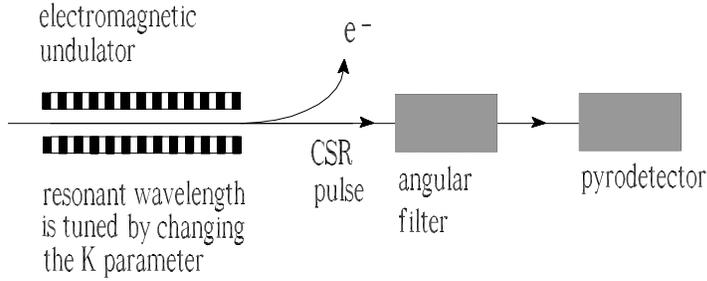,width=0.8\textwidth}
\end{center}
\caption{Scheme for electron bunch length diagnostics based on CSR
from an undulator} \label{fig:dsm16}
\end{figure}

Above, a brief treatment of undulator radiation was given in view
of beam diagnostics applications which we are going to describe in
this paragraph. Our technique relies on using an electromagnetic
undulator and on recording the energy of the coherent radiation
pulse into the central cone. The configuration is sketched in Fig.
\ref{fig:dsm16}. The electron bunch passes through the undulator
and produces a CSR pulse at some specific resonant frequency
$\omega_{0}(K)$.  The signal (energy of the CSR pulse within
central cone) is recorded by a bolometer; as we will explain in
this subsection, repetitions of this measurement with different
undulator resonant frequency allow one to reconstruct the modulus
of the bunch form factor.

The undulator has, in our scheme, a high magnetic field and a
large $K\gg 1$ value. A 40 cm magnetic period and a relativistic
factor $\gamma \simeq 10^{3}$ would produce a fundamental
wavelength of about 100 microns when $K \simeq 30$. On the other
hand, the transverse velocity of the electron is small: $\beta_{x}
\ll 1$. The combination of these conditions gives an inequality $1
\ll K \ll \gamma$ which, with a little loss of generality, we will
consider satisfied.

\begin{figure}[tb]
\begin{center}
\epsfig{file=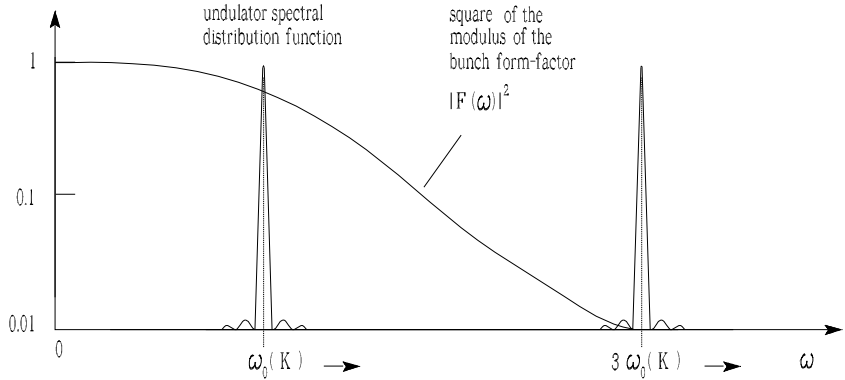,width=0.8\textwidth}
\end{center}
\caption{Some typical spectra of the longitudinal electron
distribution and undulator spectral distribution function. The
bandwidth of the undulator radiation ($\Delta\omega/\omega_{0} =
0.89/N_{\mathrm{w}}$) must always be as small as the desired
spectral resolution} \label{fig:dsm15}
\end{figure}

The principle of operation of the proposed scheme is essentially
based on the spectral properties of the undulator radiation. A
sample of undulator radiation spectrum at zero angle is shown in
Fig. \ref{fig:dsm15}.  The distribution of the radiated energy
within different harmonics depends on the value of $K$. For $K \gg
1$ we will get strong undulator maxima at $\omega = 3\omega_{0},\
5\omega_{0}$ and so forth (as already discussed, the radiation
spectrum along the undulator axis includes only odd harmonics).
The bunch form factor is the exponential function (\ref{eq:bff})
and the square modulus of bunch form-factor falls off rapidly for
wavelengths shorter than the effective bunch length.  The fast
drop off is evident and for wavelength of about three times
shorter than the effective bunch length the radiation power is
reduced to about 1 \% of the maximum pulse energy at the
fundamental harmonic.  As a result, sharp changes of the bunch
form-factor result in an attenuation of the high undulator
harmonics. This point is made clear upon examination of the
spectra presented in Fig.  \ref{fig:dsm15}: actually, the bunch
form-factor cut-off affects all higher undulator harmonics which
are, therefore, suppressed.

Our goal is to find the relationship between the expected value of
radiation energy per pulse and the square modulus of the bunch
form-factor. In general, $\mid\bar{F}(\omega)\mid^{2}$ will vary
much more slowly in $\omega$ than the sharp resonance term.  The
two functions might appear as shown in Fig.~\ref{fig:dsm15}. In
such cases, we can replace $\mid\bar{F}(\omega)\mid^{2}$ by the
constant value $\mid\bar{F}(\omega_{0})\mid^{2}$ at the center of
the sharp resonance curve and take it outside of the integral.
What remains is just the integral under the curve of Fig.
\ref{fig:dsm14}, which is, as we have seen, just equal to
(\ref{eq:cr}). Thus, the energy radiated into the central cone by
an electron bunch is given by

\begin{equation}
\Delta W_{\mathrm{CSR}} \simeq \frac{\pi
e^2A_{\mathrm{JJ}}^{2}\omega_{0}K^{2}_{\mathrm{w}}}
{c(1+ K^{2}_{\mathrm{w}}/2)}
N^{2}\mid\bar{F}(\omega_{0})\mid^{2} \ ,
\label{eq:ecc}
\end{equation}

\noindent We emphasize the following feature of this result: the
CSR radiation energy per pulse is proportional to the square
modulus of the bunch form-factor at the resonant frequency. This
fact allows one to reconstruct such quantity by repeated
measurements of $\Delta W_{\mathrm{CSR}}$ at different resonant
frequencies $\omega_0$.

\subsection{Physical remarks}

The typical textbook treatment of the undulator radiation problem
(see \cite{W}, \cite{D}) consists in finding the expression for
the undulator radiation spectrum from (\ref{eq:vf}). However, no
attention is usually paid to the region of applicability of this
expression. In general this is not correct.  In the undulator
case, it is important to realize that (\ref{eq:vf}) is valid only
when the resonance approximation is met. Naturally, the resonance
approximation cannot be expected to be valid when an insertion
device has only a few periods.  A second important limitation
concerns the far field approximation. It is generally assumed that
the distance between the undulator and the observer is much larger
that the undulator length. This approximation is often assumed
without any real justification other than the simplification that
results. Such an assumption may or may not be valid, depending on
the case.

\section{Constrained deconvolution}

If both modulus and phase of the bunch form-factor were somehow
measured, we would then known the Fourier spectrum of the bunch;
an inverse Fourier transform of the measured data
$\bar{F}(\omega)$ would then yield the desired profile function
$F(t)$ with a resolution limited by the maximum achievable
frequency range. Unfortunately, in practice, it is impossible to
extract the phase information from the CSR measurement. A more
realistic task is to measure the modulus of bunch form-factor
only: in Section 4 we proposed a method to perform such a
measurement, but independently on the technique selected, the
following problem arises, to study what information about the
bunch profile can be derived from the collected experimental data:
the problem is generally referred to as phase retrieval. There
exist some cases in which the absence of phase information is of
no consequence. For example, any symmetrical density distribution
can be recovered from the knowledge of the form factor modulus
only. An example is given by a bunch with a Gaussian density
distribution. In practice, the situations in which this assumption
is satisfied are probably rather limited.

A possible solution to the loss of phase information was suggested
by Wolf \cite{Wolf}. A step forward toward the solution of the
problem, is based on the following consideration: if

\begin{displaymath}
\bar{F}(\omega) =
\mid\bar{F}(\omega)\mid\exp\left[\I\phi(\omega)\right] \ ,
\end{displaymath}

\noindent then

\begin{displaymath}
\ln\left[\bar{F}(\omega)\right] = \ln\mid\bar{F}(\omega)\mid
+ \I\phi(\omega) \ .
\end{displaymath}

\noindent It can be proved that if
$\ln\left[\bar{F}(\omega)\right]$ is an analytical function, the
phase will be recoverable from the modulus by the relation

\begin{displaymath}
\phi(\omega) = - \frac{2\omega}{\pi}\int\limits^{\infty}_{0}
\frac{\ln\mid\bar{F}(\omega^{\prime})\mid}{(\omega^{\prime})^{2} -
\omega^{2}}\D\omega^{\prime} \ .
\end{displaymath}

\noindent Unfortunately, the analyticity of $\bar{F}(\omega)$ is
not a sufficient condition to assure analyticity for
$\ln\left[\bar{F}(\omega)\right]$ in that same region. The most
obvious reason for this is the possible existence of zeros of
$\bar{F}(\omega)$ which lead to singularities in
$\ln\left[\bar{F}(\omega)\right]$. In some situation the function
$\bar{F}(\omega)$ may have no zeros, in which case the so-called
minimum phase solution is valid \cite{Rob}. However, in general,
zeros will be present, and their localization will be unknown a
priori.

It should be noticed that, whereas the dispersion relation method
is very elegant and conceptually simple, it is not necessarily the
"simplest" method to use in practice. For any given problem, the
various possible approaches should be considered, for one may be
distinctly easier than another, depending on the problem at hand.

In the particular case of interest, namely an electron bunch
originating by the XFEL driver accelerator, we can combine
a-priori knowledge about the bunch density distribution with
measured information about the bunch form factor. We know that, in
the case under examination, the bunching process in zero order
approximation can be treated by single particle dynamical theory.
We cannot neglect the CSR induced energy spread and other wake
fields when we calculate the transverse emittance, but we can
neglect them when we calculate the longitudinal current
distribution, simply because their contribution to the
longitudinal dynamics is small. As a result, we will show that we
can define a general temporal structure of the electron bunch
after compression without any measurements.  The purpose of the
measurement is then to determine the numerical value of the
parameters on which such temporal structure depends. In this case,
sufficient information is provided by the modulus of the bunch
form-factor, allowing us to ignore the phase information. The
process of finding the best estimate of the profile function
$F(t)$ for a particular measured modulus of the bunch form-factor,
including utilization of any a priori information available about
$F(t)$ refers to a technique called "constrained deconvolution"
which will be exemplified below in detail. Although the context in
which the suggestion of this method was made first was
spectroscopy \cite{de}, the ideas apply equally well in the
present context of bunch profile reconstruction.

\begin{figure}[tb]
\epsfig{file=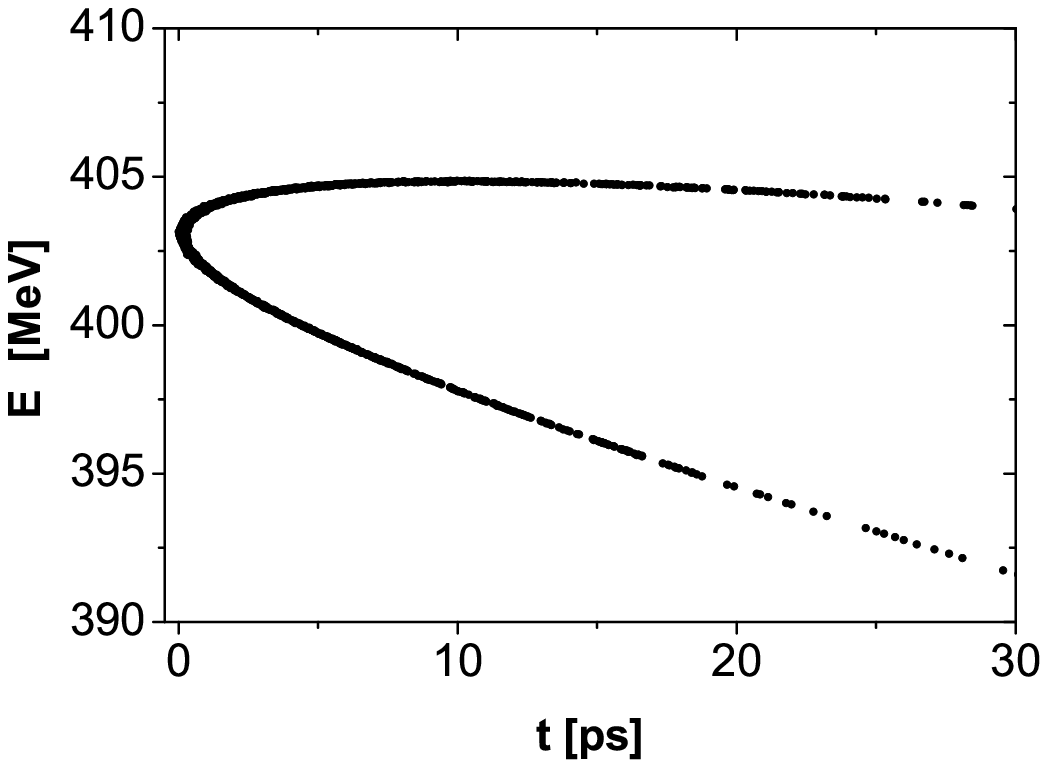,width=0.5\textwidth}

\vspace*{-53mm}

\hspace*{0.5\textwidth}
\epsfig{file=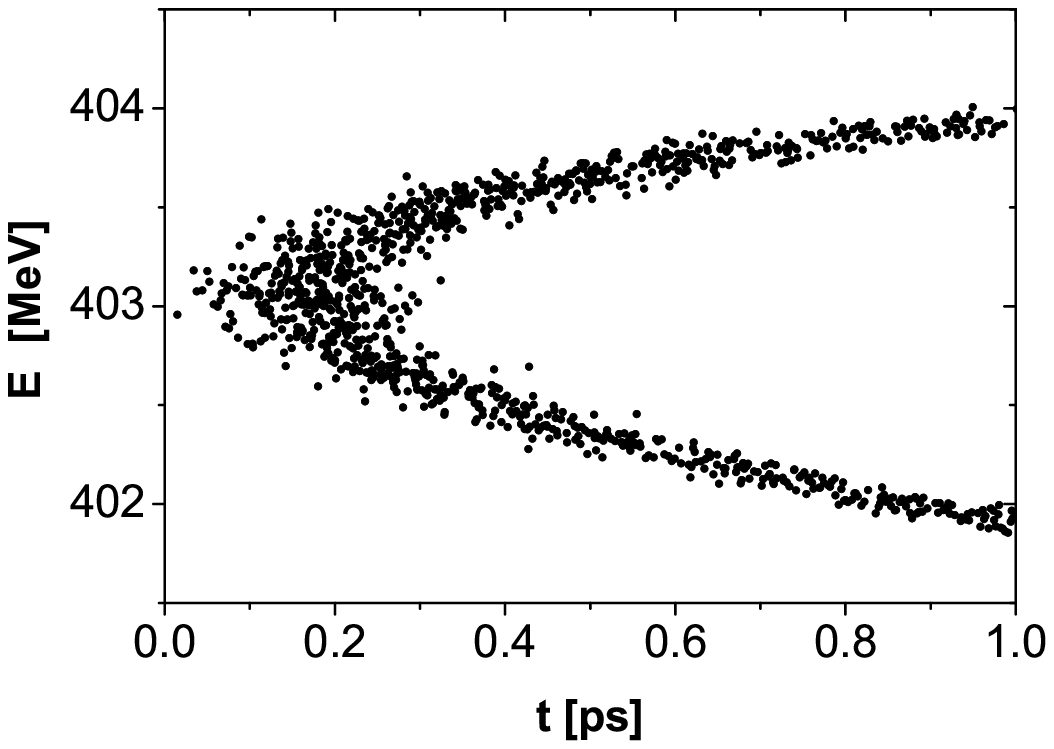,width=0.5\textwidth}

\caption{Typical phase space distribution of electrons after full
compression with a single bunch compressor. The plot on the right
presents an enlarged fraction of the plot on the left. The head of
the bunch is at the left side} \label{fig:phsp}
\end{figure}

Let us consider an example which shows the deconvolution process
in easy-to-understand circumstances. For emphasis we choose, here,
an experiment aimed at measuring the strongly non-Gaussian
electron bunch profile at TTF, Phase 2 in femtosecond mode
operation. The femtosecond mode operation is based on experience
obtained during the operation of the TTF FEL, Phase 1 \cite{ay}
and it requires one bunch compressor only. An electron bunch with
a sharp spike at the head is prepared, with rms width of about 20
microns and a peak current of about one kA. This spike in the
bunch generates FEL pulses with duration below one hundred
femtoseconds. An example of the longitudinal phase-space
distribution for a compressed beam with RF curvature effect is
shown in Fig. \ref{fig:phsp}, when the longitudinal bunch charge
distribution involves concentration of charges in a small fraction
of the bunch length.  In the femtosecond mode the first TTF
magnetic chicane only will be operational, and this will be the
default mode of operation for first lasing.

Bunch compression for relativistic situations can effectively be
achieved, at the time being, only by inducing a correlation
between longitudinal position and energy offset with an RF system
and taking advantage of the energy dependence of the path length
in a magnetic bypass section (a so-called magnetic chicane). In
order to study the CSR diagnostic for a compressed bunch, let us
first find the longitudinal charge distribution for our bunch
model when it is fully compressed by a chicane (here we assume
that the perturbation of the bunching process by CSR is
negligible). Downstream of the chicane the charge distribution is
strongly non-Gaussian with a narrow leading peak and a long tail,
as it is seen from the simulation results in Fig. \ref{fig:phsp}.
Our attention is focused on determining the longitudinal density
distribution after compression.  As first was shown in \cite{li}
there is an analytical function which provides quite a good
approximation for the simulation data.  The current distribution
along the beam after a single compression is given by

\begin{eqnarray}
& \mbox{} &
I(t) \simeq I_{0}\exp\left(-\frac{t^{2}}{2\tau_{0}^{2}}\right)  \qquad
{\mathrm{for}} \quad t_{1} > t > - \infty \ ;
\nonumber\\
& \mbox{} &
I(t) \simeq
\frac{A\exp\left(- t/\tau_{1}\right)}{\sqrt{(t+t_{0})/\tau_{1}}}
\qquad {\mathrm{for}} \quad t > t_{1} > 0 \ .
\label{eqn:u3}
\end{eqnarray}

\begin{figure}[tb]
\begin{center}
\epsfig{file=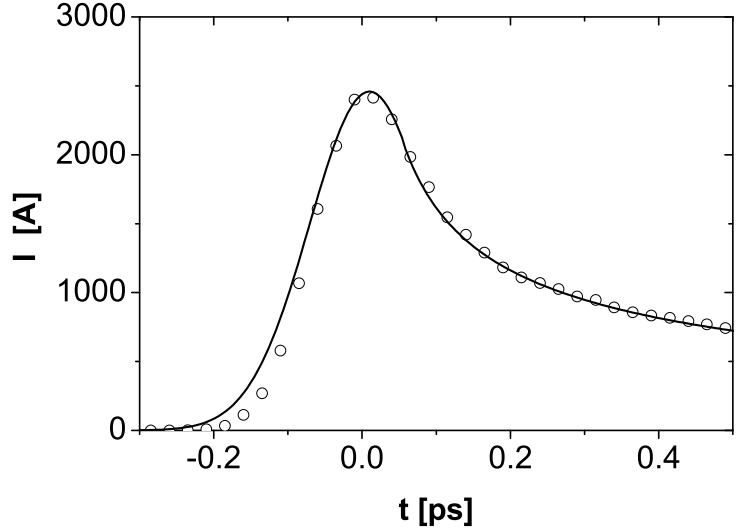,width=0.8\textwidth}
\end{center}
\caption{Current distribution along the bunch reconstructed from
an ideal set of experimental data for the form-factor and correct
value of time constant for the bunch tail (solid curve). Circles
present data from a tracking simulation code} \label{fig:i10}
\end{figure}

\begin{figure}[tb]
\begin{center}
\epsfig{file=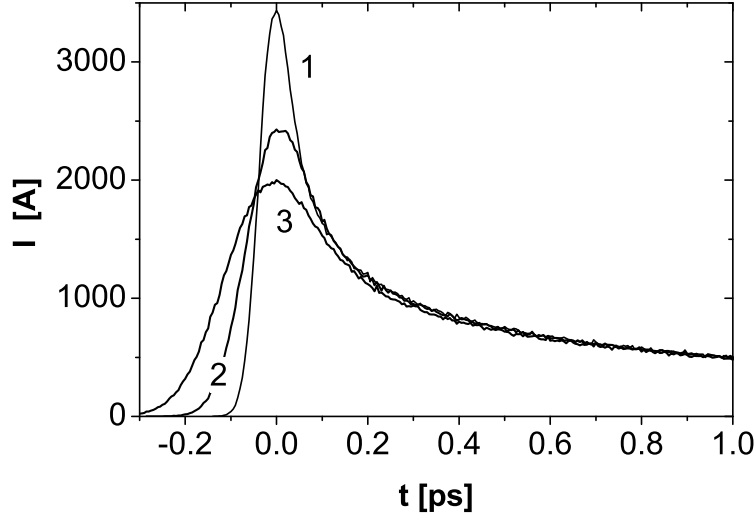,width=0.8\textwidth}
\end{center}
\caption{Typical current distribution along the bunch after full
compression with a single bunch compressor. Curve 1, 2, and 3
correspond to local energy spread of 5, 10, and 15 keV at the
entrance of the bunch compressor. The head of the bunch is at the
left side. The charge of the bunch is 3 nC} \label{fig:i5}
\end{figure}

\noindent Equation (\ref{eqn:u3}) gives a good fit. To illustrate
how good an approximation to the actual current distribution is
provided by (\ref{eqn:u3}), we show in Fig. \ref{fig:i10} the
current as a function of time, first the numerical simulation
data, then as computed from (\ref{eqn:u3}). Equation
(\ref{eqn:u3}) involves four independent parameters in the
calculation of longitudinal density distribution: $\tau_{0}$,
which characterizes the local charge concentration (non-Gaussian
feature) of the compressed density, $\tau_{1}$, which
characterizes the bunch tail and fitting parameters, $t_{0}$ and
$t_{1}$.  In this case detailed information on the longitudinal
bunch distribution (actually constants: $\tau_{0}$, $\tau_{1}$ and
$t_{0}$, $t_{1}$) can be derived from the particular measured
square modulus of the bunch form-factor by using an independently
obtained tail constant $\tau_{1}$ (for example by means of a
streak camera). A quantity of considerable physical interest is
the parameter $\tau_{0}$ which characterizes the leading peak. The
value of $\tau_{0}$ can be related to the value of the initial
local energy spread $\Delta\gamma/\gamma$ in the electron beam if
the compaction factor $R_{56}$ of the magnetic chicane is known.
Fig. \ref{fig:i5} shows plots of the peak current versus time for
several values of the local energy spread.  By comparing these
curves, some feeling can be obtained about the effective reduction
in peak current. As the energy spread is increased from 5 keV to
15 keV, the maximum peak current falls from 3.5 kA to 2 kA.

\begin{figure}[tb]
\begin{center}
\epsfig{file=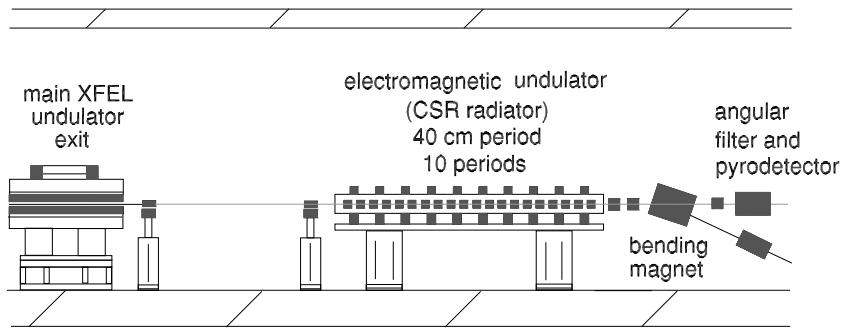,width=0.8\textwidth}
\end{center}
\caption{Bunch length measurement scheme for TTF XFEL. The CSR
source uses the spent electron beam coming from the XFEL}
\label{fig:dsm13}
\end{figure}

The highest time resolution obtainable by means of a streak camera
is in the sub-picosecond range. By means of coherent radiation,
the resolution can be made comparatively high because the
resolution is determined by the spectral range in the measurement.
Hence, this method will be useful for monitoring an ultra-short
electron bunch at TTF FEL. In the TTF FEL Phase 2 the measurement
of CSR spectrum can be made by using long period electromagnetic
undulator as shown in Fig.  \ref{fig:dsm13}. As already said, for
the complete determination of the bunch profile function,
downstream of the first TTF FEL bunch compressor, a measurement of
the bunch length can be made independently by using a
sub-picosecond resolution streak camera. The limiting resolution
of the streak camera will prevent the measurement of the spike
width. Nevertheless, the streak camera can detect the long time
tail which is outside the CSR detector response.

\begin{figure}[tb]
\begin{center}
\epsfig{file=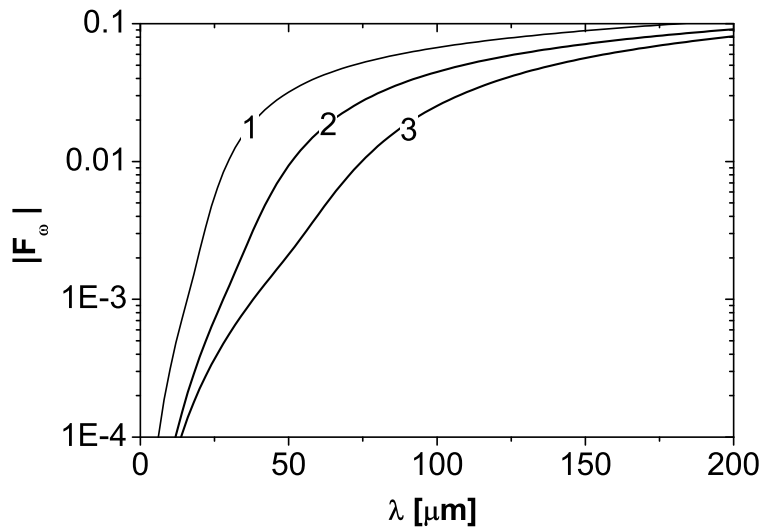,width=0.8\textwidth}
\end{center}
\caption{Form-factor of the electron bunch corresponding to the
bunch profiles shown in Fig.~\ref{fig:i5} } \label{fig:ff5}
\end{figure}

\begin{figure}[tb]
\begin{center}
\epsfig{file=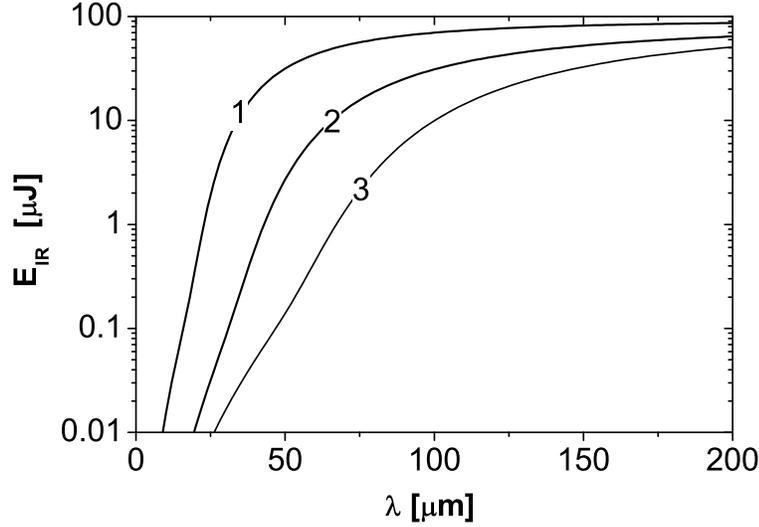,width=0.8\textwidth}
\end{center}
\caption{Energy in the radiation pulse detected within central
cone. Curves 1, 2, and 3 correspond to a local energy spread of 5,
10, and 15 keV at the entrance to the bunch compressor (see
Fig.~\ref{fig:i5}). The bunch charge is 3 nC} \label{fig:er5}
\end{figure}

The significance of the proposed scheme cannot be fully
appreciated until we determine typical values of the energy in the
radiation pulse detected within the central cone that can be
expected in practice. In the case of TTF FEL,  the bunch form
factor modulus falls off rapidly for wavelengths shorter than 40
microns. At the opposite extreme, the dependence of the form
factor on the exact shape of the electron bunch is rather weak and
can be ignored outside the wavelength range $40 \mu{\mathrm{m}} <
\lambda < 100 \mu{\mathrm{m}}$. This point is made clear upon
examination of the curves presented in Fig. \ref{fig:ff5}.  The
energy in the radiation pulse can be estimated simply as in
(\ref{eq:ecc}). From Fig. \ref{fig:er5} it is quite clear that in
the $40 \mu{\mathrm{m}} - 100 \mu{\mathrm{m}}$ wavelength region
of the CSR spectrum any local energy spread in the electron beam
smaller than 20 keV produces energies larger than $0.1
\mu{\mathrm{J}}$ in the radiation pulse.

To conclude this section we present examples of deconvolution of
computer generated data; we hope that these will provide useful
insight into the practical limits of deconvolution.  We will
demonstrate that constrained deconvolution is a useful
experimental tool testable at each step in its development.  By
far the easiest and most informative way to examine the
deconvolution process is to use a computer in place of a CSR
spectrometer. By simulating a spectrum rather than using an actual
CSR spectrometer to record it, we have a complete control over
such factors as resolution, shape and noise. Also, we know what
the perfectly deconvolved profile function would look like. In the
deconvolution of actual CSR spectral data, the presence of noise
is usually the main limiting factor: the effects of noise on
deconvolution are demonstrated in the following. For the purpose
of examining the deconvolution process, we begin with noisy data,
which of course, can be realized in a simulation process. When
other aspects of deconvolution, such as errors in the prior
information are examined, noiseless data are used.

\begin{figure}[tb]
\begin{center}
\epsfig{file=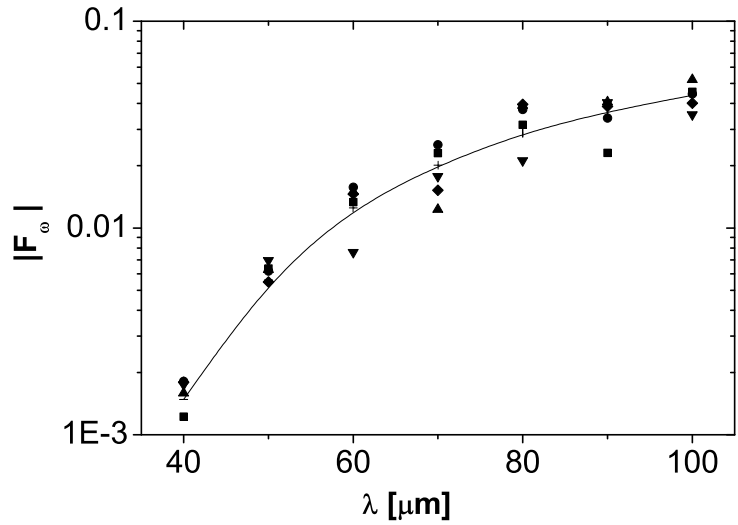,width=0.8\textwidth}
\end{center}
\caption{Electron bunch form factor corresponding to the case of
10 keV energy spread at the entrance of the bunch compressor. The
solid curve presents an ideal measurement. Symbols present five
sets of measurements with relative error $\pm 40\%$ }
\label{fig:ff40}
\end{figure}

\begin{figure}[tb]
\begin{center}
\epsfig{file=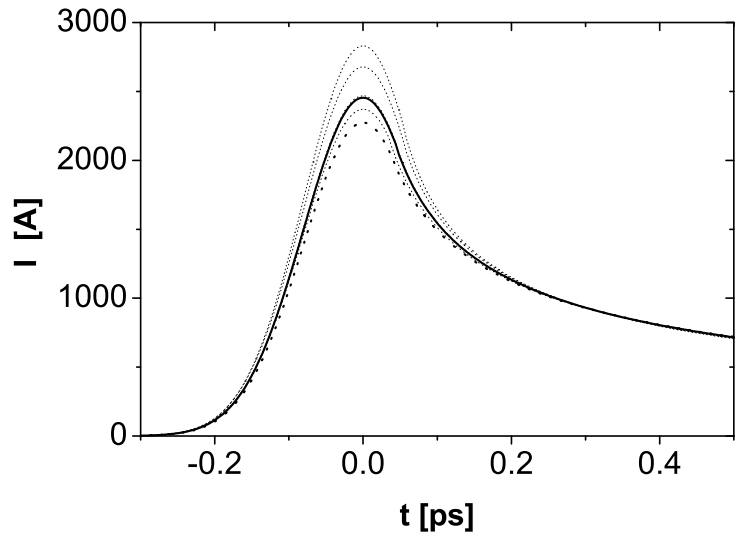,width=0.8\textwidth}
\end{center}
\caption{Current distribution reconstructed from five sets of
measurements with relative error $\pm 40\%$ (see
Fig.~\ref{fig:ff40}). The solid line corresponds to an ideal
measurement. The tail time constant in all cases is 9 ps}
\label{fig:ffer}
\end{figure}

\begin{figure}[tb]
\begin{center}
\epsfig{file=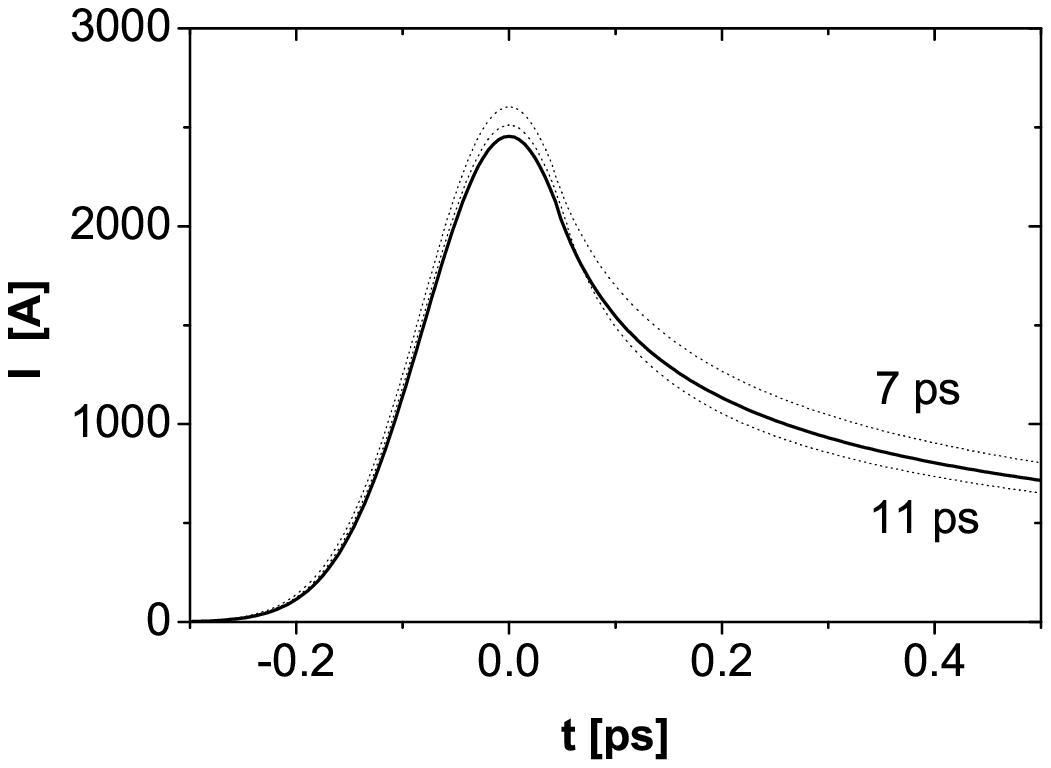,width=0.8\textwidth}
\end{center}
\caption{Current distribution reconstructed from an ideal set of
experimental data for the form-factor (see Fig.~\ref{fig:ff40}),
but with different values of time constants, 7 and 11 ps, for the
bunch tail (dotted lines). The solid line corresponds to the
correct value of the time constant, 9 ps} \label{fig:i10t}
\end{figure}

Our goal is to discover the limiting sensitivity of the
measurement technique considered. Attention is therefore turned to
the crucial question of how accurately the bunch profile
parameters can be measured by this technique when the errors in
the input data cannot be neglected. Surely, the preferable way to
answer the question about errors in bunch form-factor measurement
is the computer experiments way. The effects of various errors are
included in a full simulation of deconvolution procedure. One
important question to answer is about how well the modulus of
bunch form factor must be known in order to perform a successful
deconvolution. To obtain the sensitivity to error effects, we use
a set of input error data for a simulated bunch profile spectrum
(see Fig. \ref{fig:ff40}). A 40\% rms error (variable over the
undulator scan) is included. The sensitivities of the final bunch
profile function to the simulation input errors are summarized in
Fig. \ref{fig:ffer}. Only one random seed has been presented here
as an example, but several seeds have been run with similar
success. Fig. \ref{fig:ffer} shows that rms of CSR measurement
should be within $\pm 50 \%$.  It is seen that the constrained
deconvolution method is insensitive to the errors in the bunch
form-factor.

As shown above, the bunch profile function is given by
(\ref{eqn:u3}). An important issue to deal with consists in
determining the dependence of the deconvolved function on the tail
constant. One way to investigate this matter is to use a set of
tail constants which differ from the real one in the deconvolution
of a simulated spectrum. Fig. \ref{fig:i10t} shows the results of
such a test. At first glance all three traces of Fig.
\ref{fig:i10t} are practically identical, indicating that the tail
parameter is less important than it would have probably been
expected. We can therefore conclude that the width of the tail
parameter should be within $\pm 50\%$ of the proper value.
However, if the correct tail function is not readily determined
(by a streak camera) there is some justification for deconvolution
with a "good guess" because even the extreme case shown in Fig.
\ref{fig:i10t} is superior to trace the profile function without
prior information.

\section{Concluding remarks}

Electron pulse-shape measurement is achieved by a three-step
process.

\noindent (i)  In the first step spectrum of coherent radiation
$P(\omega)$ is measured. Coherent radiation from dipole magnet is
emitted over a wide spectrum and should be analyzed with a
spectrometer or interferometer. The proposed technique of using
CSR from an undulator may be a more promising approach.  The basic
measurement in this case is extremely simple:  the energy of the
coherent radiation pulse within the central cone is recorded by a
bolometer.  In order to use the radiation from the undulator it
has to be filtered: for such a purpose, a simple pinhole can be
used.

\noindent (ii) The second step is to deduce the modulus of the
bunch form-factor from the equation $P(\omega) =
p(\omega)N^{2}\mid\bar{F}(\omega)\mid^{2}$; $p(\omega)$  should be
applied to the result of the measurement, to obtain the bunch form
factor.  A careful analysis of the diagnostic instrumentation is
required to find its exact $p(\omega)$ function. The computation
of the single-particle spectrum  for practical accelerator
applications is a rather difficult problem. The long wavelength
synchrotron radiation from bending magnets in the near zone
integrates over many different vacuum chamber pieces and the
quantity $p(\omega)$ is usually very difficult to know with good
accuracy within the long wavelength range. In principle, modern
computers allow one to perform direct calculation of the  factor
$p(\omega)$. The results of such simulations will depend on a
large number of parameters.  They will provide the possibility of
obtaining a numerical answer for a specific set of input data, but
hardly help to understand the physics of the problem.  Success
come to those who start from the physical point of view and begin
by making the right kind of approximations, knowing what is large
and what is small in a given complicated situation. We hope that
the simple model employed here offers a useful introduction to the
more formidable problems which arise in the computation of
spontaneous emission spectrum distortions for real accelerators.

\noindent (iii) The third step consists in the reconstruction of
the electron bunch profile from the measured data. The CSR
spectrum measurement has an important limitation:  it does not
provide any information on the phase content of the pulse. It is
clear that the retrievability of the information about the
electron bunch profile function will in general not be complete,
for it is the particular measured square modulus of the
form-factor that can be obtained, not the complex form-factor
itself. No spectral phase information is available, and the
complete profile function cannot be obtained in general.
Nevertheless, the less complete information can be exceedingly
useful in many cases. For example, downstream of the bunch
compressor the bunch charge distribution is strongly non-Gaussian
with a narrow leading peak and a long tail. In this case detailed
information about the longitudinal bunch distribution can be
derived from the measured square modulus
$\mid\bar{F}(\omega)\mid^{2}$ by measuring the bunch tail constant
with a streak camera and by using all a-priori available
information about profile function. In this paper we demonstrate
that constrained deconvolution is a useful experimental tool
testable at each step in its development. Although the present
work is concerned primarily with fully compressed electron bunch
downstream of the first bunch compressor, its applicability is not
restricted to that area only: for example, multistage bunch
compression is a suitable candidate for treatment by the methods
described here.

\section*{Acknowledgments}

We thank J. Botman, W. Brefeld, R. Brinkmann, B. Faatz, A. Fateev,
J. Feldhaus, M. Koerfer, O. Kozlov, J. Krzywinski, J. Luiten, T.
Moeller, D. Noelle, J. Pflueger, P. Piot, E. Ploenjes, J.
Rossbach, S. Schreiber, M. v.d. Wiel for many useful discussions.
We thank J.R.~Schneider and D.~Trines for interest in this work.

\end{document}